\def\mdot{\dot M}

\documentclass[longnamesfirst]{emulateapj}
\usepackage{natbib}
\usepackage{apjfonts,lscape}
\tighten

\received{2007 November 14}
\accepted{2008 April 21}

\slugcomment{The Astrophysical Journal, in press}
\shortauthors{BAUER ET AL.}
\shorttitle{SN\,1996\MakeLowercase{cr}: SN\,1987A's Wild Cousin?}

\begin{document}

\title{Supernova 1996\MakeLowercase{cr}: \hbox{SN\,1987A}'s Wild Cousin?}

\author{
F.~E.~Bauer,\altaffilmark{1} 
V.~V.~Dwarkadas,\altaffilmark{2}
W.~N.~Brandt,\altaffilmark{3}
S.~Immler,\altaffilmark{4} 
S.~Smartt,\altaffilmark{5}
N.~Bartel,\altaffilmark{6}
and 
M.~F.~Bietenholz\altaffilmark{6,7}
}

\altaffiltext{1}{{\it Chandra} Fellow, Columbia Astrophysics Laboratory, 550 W. 120th St., Columbia University, New York, NY 10027; feb@astro.columbia.edu}
\altaffiltext{2}{Department of Astronomy and Astrophysics, University of Chicago, 5640 South Ellis Avenue, RI 451, Chicago, IL 60637}
\altaffiltext{3}{Department of Astronomy \& Astrophysics, The Pennsylvania State University, 525 Davey Lab, University Park, PA 16802.}
\altaffiltext{4}{Goddard Space Flight Center, Code 662, Greenbelt, MD 20771, USA}
\altaffiltext{5}{Astrophysics Research Centre, School of Maths and Physics, Queen's University Belfast, Belfast, BT7 1NN Northern Ireland, UK}
\altaffiltext{6}{Department of Physics and Astronomy, York University, Toronto, ON M3J 1P3, Canada}
\altaffiltext{7}{Hartebeesthoek Radio Observatory, PO Box 443, Krugersdorp, 1740, South Africa}

\begin{abstract}
We report on new VLT optical spectroscopic and multi-wavelength
archival observations of \hbox{SN\,1996cr}, a previously identified
ultraluminous \hbox{X-ray} source known as Circinus Galaxy X-2. The
spectrum of the optical counterpart confirms \hbox{SN\,1996cr} as a
{\it bona fide} type~IIn SN, in accordance with its tentative SN
classification at \hbox{X-ray} wavelengths. \hbox{SN\,1996cr} is one
of the closest SNe ($\approx$3.8~Mpc) in the last several decades and
in terms of flux ranks among the brightest radio and \hbox{X-ray} SNe
ever detected. Optical imaging from the Anglo-Australian Telescope
archive allows us to isolate the explosion date to between 1995-02-28
and 1996-03-16, while the wealth of optical,
\hbox{X-ray}, and radio observations that exist for this source
provide relatively detailed constraints on its post-explosion
expansion and progenitor history, including an preliminary angular
size constraint from VLBI. Archival \hbox{X-ray} and radio data imply
that the progenitor of \hbox{SN\,1996cr} evacuated a large cavity just
prior to exploding via either a sped-up wind or a pre-SN explosion.
The blast wave likely spent $\sim$1--2~yrs in relatively uninhibited
expansion before eventually striking the dense circumstellar material
which surrounds \hbox{SN\,1996cr} to become a prodigious \hbox{X-ray}
and radio emitter. The \hbox{X-ray} and radio emission, which trace
the progenitor mass-loss rate, have respectively risen by a factor of
$\ga$2 and remained roughly constant over the past $\approx7$~yr. This
behavior is reminiscent of the late rise of \hbox{SN\,1987A}, but
three orders of magnitude more luminous and much more rapid to
onset. \hbox{SN\,1996cr} may likewise provide us with a younger
example of \hbox{SN\,1978K} and \hbox{SN\,1979C}, both of which
exhibit flat X-ray evolution at late times.  The optical spectrum
suggests that the progenitor was a massive star that shed some of its
outer envelope (many M$_{\odot}$) prior to explosion, while the
complex Oxygen line emission hints at a possible concentric shell or
ring-like structure.  Taken together, this implies that a substantial
fraction of the closest SNe observed in the last several decades have
occurred in wind-blown bubbles, and argues for the phenomena being
widespread.
\end{abstract}


\section{Introduction}\label{intro}

Although thousands of supernovae (SNe) have been discovered to date,
only several dozen have been detected at
either \hbox{X-ray}\footnote{See
http://lheawww.gsfc.nasa.gov/users/immler/supernovae\_list.html for a
complete list of X-ray SNe and references} or radio\footnote{See
http://rsd-www.nrl.navy.mil/7213/weiler/kwdata/RSNtable.txt for a list
of radio SNe and references} wavelengths (designated XSNe or RSNe,
respectively). Many of these XSNe and RSNe were originally detected
simply as variable
\hbox{X-ray} or radio sources, and only verified as true SNe after
careful examination of archival optical data or through optical
follow-up \citep[e.g.,][]{Rupen1987, Ryder1993}. This particular path
to discovery has meant that only a handful of these sources have been
well-studied during the first several hundred days, a period which is
critical for identifying and characterizing the true nature of the
SNe.\footnote{By contrast, XSNe and RSNe do comprise a substantial
fraction of all SNe well-studied beyond a few hundred days, as such
SNe end up being nearby and remain relatively bright for years to
decades.} Such SNe typically turn out to be core-collapse SNe of
either type Ibc, which have been associated with long-duration
gamma-ray bursts (GRBs), or type~II, whereby the intense \hbox{X-ray}
and/or radio emission is thought to stem from the interaction between
the expanding shock and a dense progenitor wind. When well-sampled,
the additional multi-wavelength constraints from XSNe and RSNe can
provide physical insights into the late evolutionary stages of massive
stars that are otherwise impossible to obtain. For instance, while
robust constraints on the overall time-averaged mass-loss in various
phases of stellar evolution exist, there has been a long-standing
debate over the number and sequence of various evolutionary stages for
massive stars, and relatively few observational constraints on the
actual evolution of mass-loss within many particular
stages \citep[e.g.,][]{Lamers1991, Langer1994, Stothers1996,
Maeder2000}. Because the SN blast wave travels orders of magnitude
faster than the stellar wind, studying the interaction between the
blast wave and the progenitor stellar wind allows us to probe
tens of thousands of years of evolution in a matter of decades. SNe
progenitors are considered to provide the bulk of all processed
stellar material additionally, and thus the characterization of these
sources can likewise help further our understanding of overall
galactic chemical enrichment.

Here we report on the spectroscopic confirmation of one such
serendipitous source, \hbox{SN\,1996cr} \citep{Bauer2007}, and present
a multi-wavelength follow-up study using archival data to determine its
explosion date and temporal properties. \hbox{SN\,1996cr} was
originally detected as Circinus Galaxy (CG) X-2, an ultraluminous
X-ray source in the nearby Circinus Galaxy \citep{Freeman1977}, which
\citet{Bauer2001} found to exhibit many characteristics of a young,
type~II SN enshrouded in a dense circumstellar environment.  In
particular, it demonstrated a factor of $>$30 increase in \hbox{X-ray}
flux between \hbox{1997--2000}, a $kT\sim$10~keV thermal spectrum with
a strong, blended Fe emission-line component at 6.85~keV, and spatial
coincidence with a strong radio and H$\alpha$-emitting point
source. The proximity of this SN affords us a rare opportunity to
study in detail a type~II SN which is strongly interacting with its
circumstellar medium.

This paper is organized as follows: 
data and reduction methods are detailed in $\S$\ref{data}; 
confirmation of \hbox{SN\,1996cr} as a SN and isolation of its
explosion date are provided in $\S$\ref{confirm};
overall temporal and spectroscopic constraints for \hbox{SN\,1996cr}
are investigated in $\S$\ref{character};
and finally conclusions and future prospects are explored in
$\S$\ref{conclude}.
Throughout this paper, we adopt a distance of $3.8\pm0.8$~Mpc to the
Circinus Galaxy \citep[converted from][]{Freeman1977}. While the
Circinus Galaxy lies close to the Galactic Plane ($|b|=$3\fdg8), it is
located within a Galactic ``window'' with a visual extinction of
$A_{V}=1.5\pm0.2$ and a neutral hydrogen column density of $N_{\rm
H}=(3.0\pm0.3)\times10^{21}$ cm$^{-2}$ \citep[whereas neighboring
regions typically have $A_{V}=3.0$ and $N_{\rm
H}=(5$--$10)\times10^{21}$~cm$^{-2}$;][]{Schlegel1998, Dickey1990}.
Due to its $\sim65\arcdeg$ inclination, however, there is significant
internal obscuration as well [\citealp[$N_{\rm H}\sim(5$--$8)\times10^{21}$
cm$^{-2}$ typically;][]{Bauer2001}].

\section{Observational Data and Reduction Methods}\label{data}

We describe below the extensive observational data used to constrain
the properties of \hbox{SN\,1996cr}. Fortunately, plentiful archival
data exist for the Circinus Galaxy due to the fact that it hosts the
second closest Compton-thick active galactic nuclei (AGN) to our own
Galaxy and exhibits signs of vigorous star
formation \citep[e.g.,][]{Matt1999}. The resulting multiwavelength
dataset provides good constraints on the explosion date and yields
useful long-term \hbox{X-ray} and radio light curves. \hbox{SN\,1996cr}
lies 25\arcsec to the south of the Circinus Galaxy nucleus. We adopt
the position of \hbox{$\alpha=14^{\rm h}13^{m}10$\fs01
$\delta=-65\arcdeg20\arcmin44\farcs4$} (J2000), determined from the
radio observations.

\subsection{VLT Spectroscopy}\label{data_spectra}

\begin{figure*}
\vspace{-0.1in}
\centerline{
\hglue-1cm{\includegraphics[width=18.0cm]{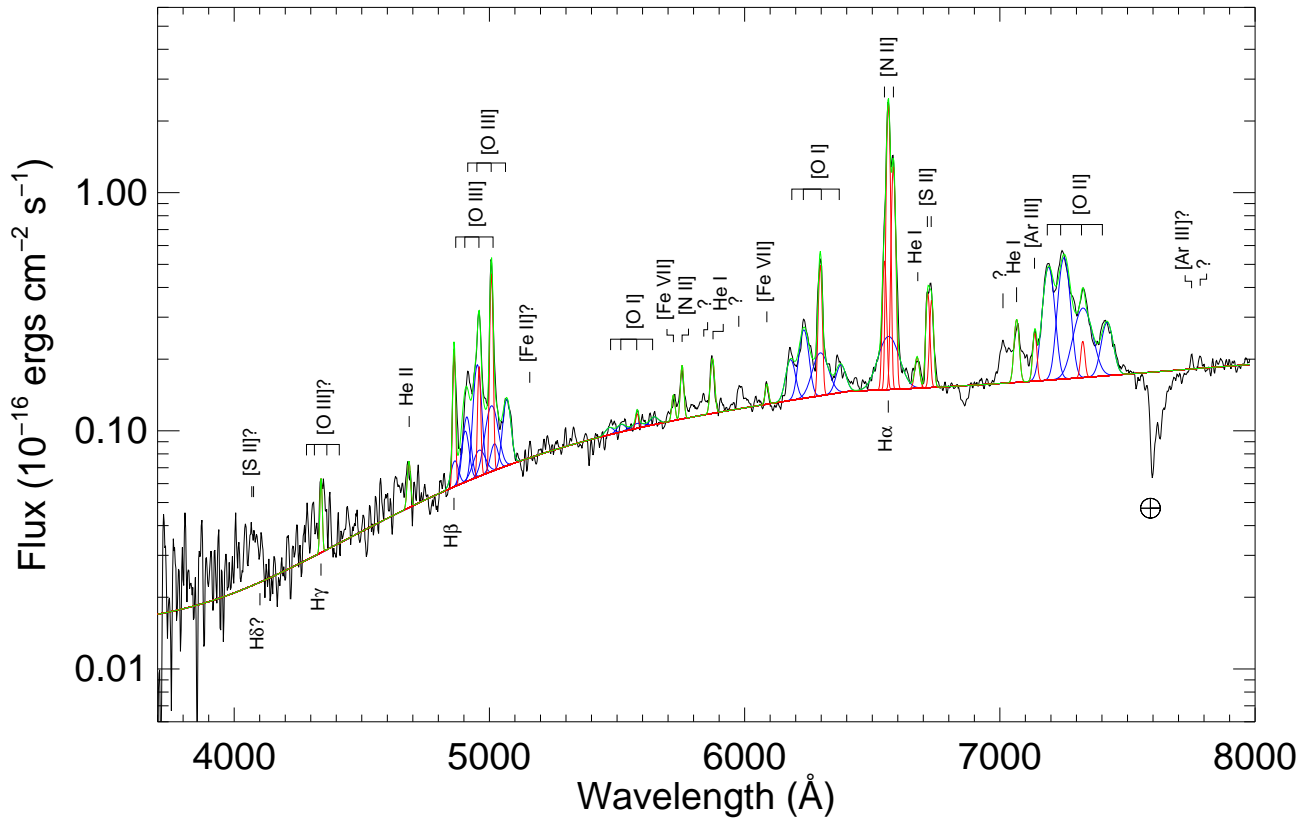}}
}
\vspace{-0.2cm} 
\figcaption[sn1996cr_spec.eps]{
%
%
Flux-calibrated VLT FORS spectrum of \hbox{SN\,1996cr} from 2006
January 26. The spectrum suffers from an extinction of
$A_{V}\sim4$--6, which we have not corrected. Obvious emission lines
have been identified. There is strong, narrow H$\alpha$ emission
(FWHM$\approx714$~km~s$^{-1}$) as well as several broad O complexes
showing a distinct underlying velocity structure. An empirical model
has been fit to the data as described in
$\S$~\ref{opt_spectra_results}; the green line shows the overall fit to
the spectrum while the red (blue) lines show the individual narrow
(broad) components. Strong, broad unidentified residuals remain around
the \ion{He}{1} lines ($\lambda\lambda$5876,7065); this emission is
likely associated with the structure of a He shell or the progenitor
wind, but is not easily disentangled.
%
%
%
\label{fig:sn1996cr-spec}}
\vspace{0.2cm} 
\end{figure*}

SN\,1996cr was visited several times in service mode with the VLT
FORS~I spectrograph in two separate campaigns. The first program was
initiated over the course of 2005-03-06 through 2005-03-17 but never
completed, resulting in a low signal-to-noise spectrum. As such, we do
not provide any further details.  The second program was executed over
three nights (2006-01-26, 2006-02-02, and 2006-03-10), yielding a
high-quality spectrum confirming the ambiguous features seen in the
spectrum from our first program. Our analyses focus only on the
spectrum from the first night, as the observations on the subsequent
two nights suffer from significantly worse seeing
($\approx$1\farcs5--2\farcs0) such that the co-addition of these
frames failed to improve the signal-to-noise of the final
spectrum. Our observations on the first night consisted of four 1200~s
exposures taken with the 300V grism and the TEK (24$\mu$ pixel) CCD
using a slitwidth of 2\farcs0. The grism and CCD combination provided
a dispersion of 2.66\,\AA~pixel$^{-1}$ and a total useful wavelength
coverage of 3600--7990\,\AA. These images were taken in excellent
seeing conditions, in which stellar sources had a full-width
half-maximum (FWHM) measured at 0\farcs6--0\farcs7 on the FORS~I chip
over the course of the night. Our target, \hbox{SN\,1996cr}, had an
$\approx20$\% larger FWHM image size than the point sources in the
acquisition image, suggesting that it may be spatially extended or
possibly contaminated by a coincident H\,{\sc ii} region. We
investigated this further by measuring the FWHM orthogonal to the
dispersion axis for the strong emission lines
[\ion{O}{3}]$\lambda$5007 and H$\alpha$, which had sizes of 0\farcs74
and 0\farcs85, respectively. The inherent FWHM slowly increased with
decreasing wavelength, such that the FWHMs quoted above were 5$\pm$3\%
and 35$\pm$4\% larger than their stellar equivalents,
respectively. Thus the [\ion{O}{3}]$\lambda$5007 FWHM is formally
unresolved, while the H$\alpha$ line is extended. The strength, extent,
and visible asymmetry in the FWHM profile of the extended H$\alpha$
are fully consistent with the underlying H\,{\sc ii} region (see
$\S$\ref{confirm} for details), although we caution that
marginally-resolved light echoes could potentially contaminate and
broaden the PSF \citep[e.g.,][]{Sugerman2003}. The spectral resolution
(11\,\AA, or 500\,km~s$^{-1}$) was estimated from the width of the
arclines taken with a 1\farcs0 slit, which matched closely the on-sky
image size of \hbox{SN\,1996cr}.  After bias subtraction and
flat-fielding using standard techniques, the co-added spectra were
extracted, wavelength calibrated, and flux calibrated (using the
standard LTT4816). Fig.~\ref{fig:sn1996cr-spec} shows the resulting
optical spectrum from 2006-01-26. We conservatively estimate the flux
errors on the spectrum at 20\%.

\subsection{Archival Optical Data}\label{data_optical}

We searched through the European Southern Observatory (ESO), Anglo
Australian Telescope (AAT), United Kingdom Schmidt Telescope (UKST),
and {\it Hubble Space Telescope} ({\it HST}) archives. We only list
here data that ultimately were used to constrain the explosion date of
\hbox{SN\,1996cr}. Relevant data are listed in
Tables~\ref{tab:data_eso}--\ref{tab:data_ukst}. Aperture photometry
was performed throughout, with PSF-fitting employed as a cross-check;
there is only one instance where the PSF-fitting magnitude differed
from the aperture-measured value at $\ga1\sigma$, which we explicitly
document below.

\subsubsection{ESO}\label{data_eso}

\begin{figure*}
\vspace{-1.7in}
\centerline{
\includegraphics[height=6.0in]{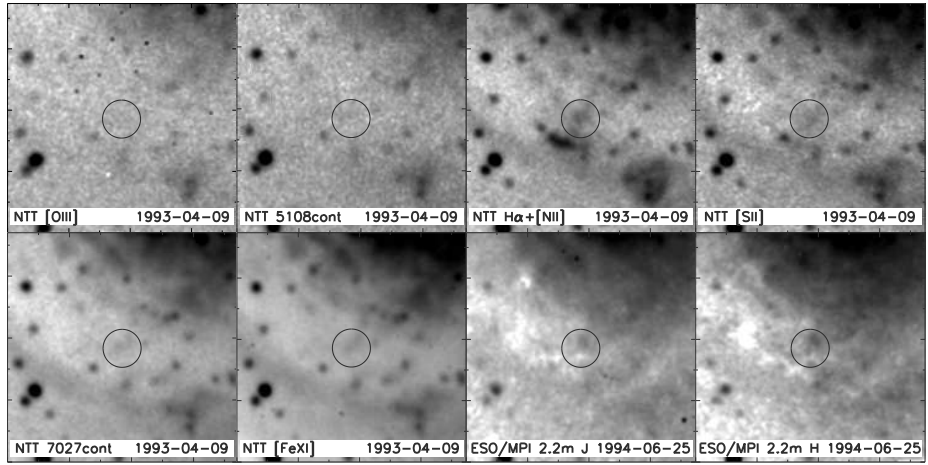}
}
\vspace{-1.7in} 
\figcaption[NTT.eps]{
Postage-stamp images centered on \hbox{SN\,1996cr} from ESO imaging
observations in 1993--1994 courtesy of E. Oliva and A. Marconi
(private communication, 2007). Images are $24\arcsec$ on a side. The
position of \hbox{SN\,1996cr} is denoted by a $2\arcsec$ radius
circle. Each image is labeled, denoting both the instrument and
filter used to acquire it and the observation date.
\label{fig:eso_images}}
\vspace{0.1in} 
\end{figure*}

The Circinus Galaxy was observed extensively with SUSI at the NTT on
1993 April 9 using several narrow-band filters (\#369:[\ion{O}{3}],
\#430:5100\,\AA~continuum, \#629:H$\alpha$+[\ion{N}{2}],
\#700:[\ion{S}{2}], \#443:7000\,\AA~continuum,
\#415:[\ion{Fe}{11}]$\lambda$7892)\footnote{\url{http://filters.ls.eso.org/efs/efs\_fi.htm}} as well as with the IRAC2
camera at the ESO/MPI 2.2m telescope on 1994 June 25 using broad-band
$J$ and $H$ filters. Reduced images were kindly provided by E. Oliva
and A. Marconi (private communication, 2007), for which cutouts of
\hbox{SN\,1996cr} are shown in Fig.~\ref{fig:eso_images}. Details of
the observations and reduction procedures are given in
\citet{Marconi1994}. The [\ion{O}{3}], H$\alpha$+[\ion{N}{2}], and
[\ion{S}{2}] images were the only images provided to us in a
flux-calibrated state, and therefore are the only ones for which we
measure photometry. Magnitudes for the region in the vicinity of
\hbox{SN\,1996cr} were measured using a 1\farcs0 radius circular
aperture and previously established zero points. An aperture
correction of 0.2~mags was estimated empirically using several bright,
isolated point sources in the images. The resulting aperture-corrected
magnitudes are presented in Table~\ref{tab:data_eso}.

\subsubsection{{\it HST}}\label{data_hst}

\begin{figure*}
\vspace{-1.8in} 
\centerline{
\includegraphics[height=6.0in]{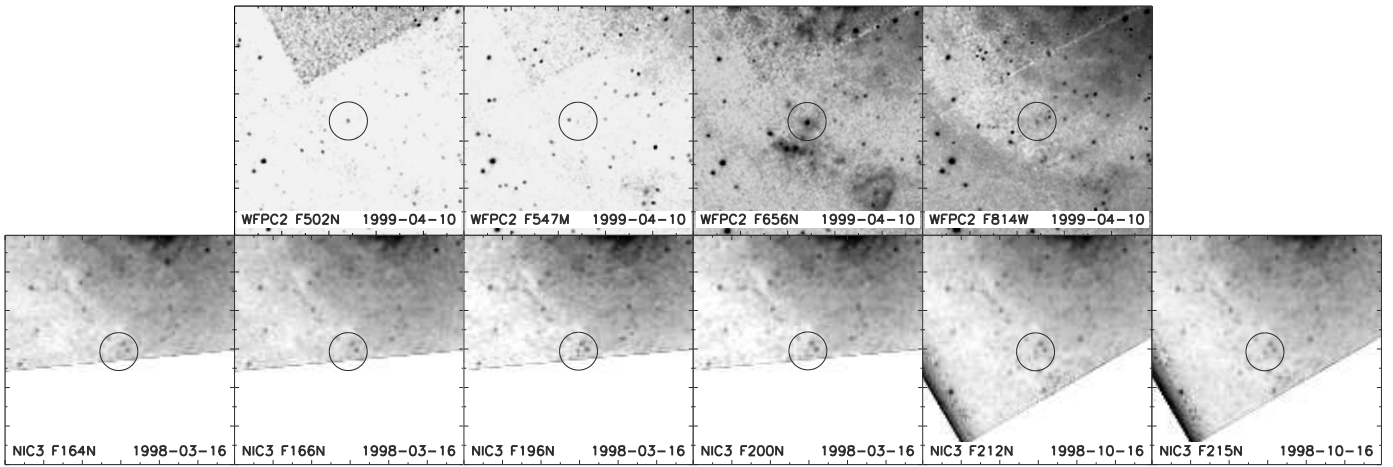}
}
\vspace{-1.7in} 
\figcaption[HST.eps]{
Postage-stamp images centered on \hbox{SN\,1996cr} from {\it HST} WFPC2
and NIC3 imaging observations spanning 1996 to 1999. \hbox{SN\,1996cr}
appears as a strong, unresolved emission-line source in the F502N and
F656N images, but displays only weak continuum as evidenced its
unremarkable appearance in all of the other {\it HST} images and the
fainter broad-band magnitudes in Table~\ref{tab:data_hst}. See
Fig.~\ref{fig:eso_images} caption for details of image properties.
\label{fig:hst_images}}
\vspace{0.1in} 
\end{figure*}

The Circinus Galaxy was observed with both the WFPC2 and NICMOS
instruments aboard {\it HST} on four separate occasions, as outlined
in Table~\ref{tab:data_hst}.  Details of the photometric and
astrometric reduction of the WFPC2 data are given in
\citet{Bauer2001}. Briefly, after standard calibration of the {\it
HST} images, we used the {\sc iraf} package {\sc daophot} to measure
aperture-corrected magnitudes using a 0\farcs2 radius aperture for all
sources down to the \hbox{2$\sigma$} detection limit. The {\it HST}
images were then aligned to the {\it Hipparcos}/Tycho astrometric
reference frame to $\approx$0\farcs4. Due to the negligible overlap of
the F606W observation with the other filters and complete lack of
coverage of \hbox{SN\,1996cr} itself, we do not discuss it further.
For the NICMOS data, of which only the NIC3 images provided useful
imaging, we used the standard pipeline data products. We performed
photometry using a 0\farcs5 radius aperture and applied band-dependent
aperture corrections determined from a set of isolated, unsaturated
point sources in the vicinity of \hbox{SN\,1996cr}. Cutout images
of \hbox{SN\,1996cr} are presented in
Fig.~\ref{fig:hst_images}. Notably, \hbox{SN\,1996cr} appears to lie at
the center of a diffuse patch of H$\alpha$ emission in the F656N image
and our aperture magnitude here differs from a PSF-fitting one at
$\ga3\sigma$ (0.11 magnitudes). We thus adopt the PSF-fitting
magnitude, as well as an additional systematic error of 0.1 magnitudes
to reflect the larger uncertainty associated with the deblending.

\begin{figure*}
\vspace{-1.8in} 
\centerline{
\includegraphics[height=6.0in]{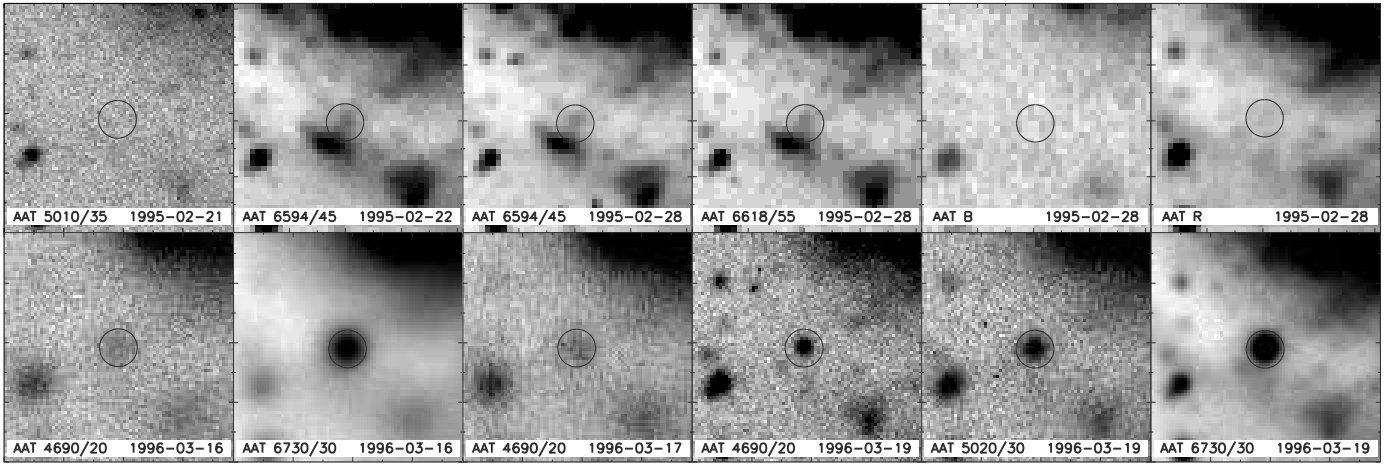}
}
\vspace{-1.7in} 
\figcaption[AAT.eps]{
Postage-stamp images centered on \hbox{SN\,1996cr} from AAT imaging
observations. The top six images from 1995-02-21 to 1995-02-28 represent
the comparison epoch, while the bottom six images from 1996-03-16 to 
1996-03-19 represent the discovery epoch. Despite the poor quality of some
images, it is clear that \hbox{SN\,1996cr} is consistently luminous in
all discovery images spanning several days. See
Fig.~\ref{fig:eso_images} caption for details of image properties.
\label{fig:aat_images}}
\vspace{0.1in} 
\end{figure*} 

\begin{figure}
\vspace{-1.7in}
\centerline{
\includegraphics[height=6.0in]{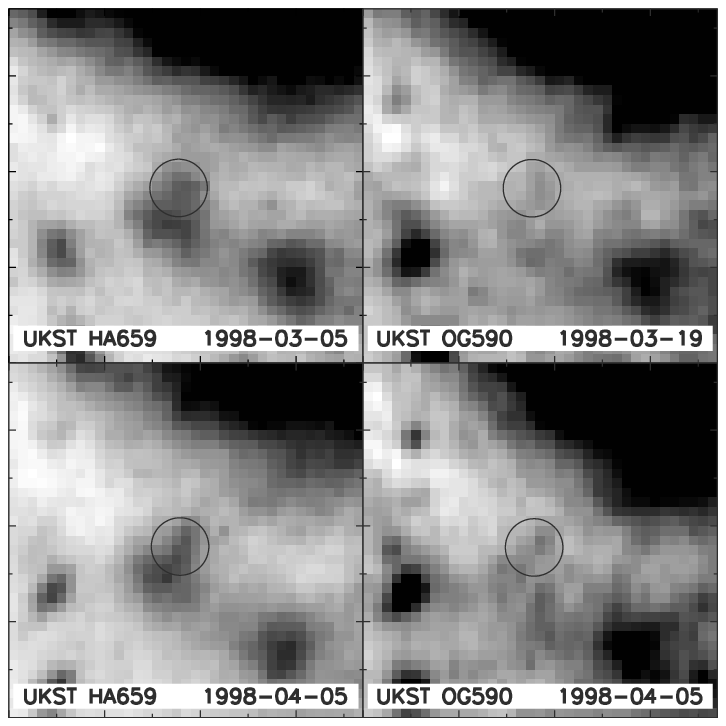}
}
\vspace{-1.7in}
\figcaption[UKST.eps]{
Postage-stamp images centered on \hbox{SN\,1996cr} from UKST imaging
observations in 1998. \hbox{SN\,1996cr} is still evident as a
slight excess in the H$\alpha$ images compared to similar narrow-band
images in Figs.~\ref{fig:eso_images} and \ref{fig:aat_images}. See
Fig.~\ref{fig:eso_images} caption for details of image properties.
\label{fig:ukst_images}}
\vspace{0.1in}
\end{figure} 

\subsubsection{AAT}\label{data_aat}

The Circinus Galaxy was observed numerous times with the TAURUS
Fabry-Perot instrument on the AAT during 1995-02-21 to 1995-02-28 and
1996-03-15 to 1996-03-20. The data were retrieved through the AAT
archive.\footnote{\url{http://site.aao.gov.au/arc-bin/wdb/aat\_database/observation\_log/make}}
The observations were comprised primarily of narrow-band imaging and
TAURUS spectral imaging cubes centered on the
\ion{He}{2}$\lambda$4686, [\ion{O}{3}]$\lambda\lambda$5007,4959,
H$\alpha+$[\ion{N}{2}]$\lambda\lambda$6583,6548, and
[\ion{S}{2}]$\lambda\lambda$6731,6716 emission lines. The original
investigators discuss their observational design
in \citet{Veilleux1997} and \citet{Elmouttie1998}, while more complete
data reduction procedures are outlined in \citet{Gordon2000}.  To
summarize, the TAURUS instrument was used in the angstrom imaging
mode, wherein narrow-band filters were used at different tilt angles
to isolate the lines of
\ion{He}{2}, [\ion{O}{3}], H$\alpha+$[\ion{N}{2}], and [\ion{S}{2}]
\citep{Bland-Hawthorn1998}.  Each square pixel subtended 0\farcs315 on the
sky and the atmospheric seeing at FWHM averaged
$\approx$1\farcs2--2\farcs0. Unfortunately, a significant portion of
the archived TAURUS images for the Circinus Galaxy have only limited
accompanying CCD calibration data; there were often no obvious
superbias frames, dark frames, dome-flats, or flux standards, so the
photometric quality of the reduced data is limited.

Our best-effort reduction proceeded as follows. The overscan regions
were used to subtract the bias from each frame and each image was
flat-fielded using a sky flat. When multiple frames in a given filter
were available, the images were combined to reject cosmic rays using
the {\sc iraf} task {\sc crrej}. Source detection and photometry were
performed with SExtractor \citep{Bertin1996}. A set of 40 relatively
bright, isolated, unsaturated stars, which are detected in all the AAT
and {\it HST} images, were chosen to register the AAT images to the
{\it HST} coordinate frame. The alignment of each AAT image to this
reference frame is accurate to $\approx$0\farcs1--0\farcs2
(\hbox{1$\sigma$}). The zeropoints for the narrow-band images were obtained by
bootstrapping the AAT magnitudes to the well-determined {\it HST}
ones. To this end, {\it HST} colors were obtained for the 40 stars and
matched to standard main sequence stellar templates from the Bruzual
Atlas,\footnote{
\url{http://www.stsci.edu/hst/observatory/cdbs/bz77.html}} modified by a
standard Galactic dust model with $E(B-V)=1$ \citep[the extinction
measured toward the Circinus Galaxy; e.g.,][]{Schlegel1998} using the
{\sc iraf} package {\sc synphot}. Among the 40 calibration stars, 25
had colors consistent with one of the stellar templates. These
particular stellar templates and the TAURUS-specific transmission
curves\footnote{\url{http://www.aao.gov.au/local/www/cgt/ccdimguide/filtercat.html}}
were input into {\sc synphot} to convert between {\it HST} and TAURUS
magnitudes. We used the {\sc iraf} task {\sc fitparams} to compare
interactively the {\it HST}-derived magnitudes to the SExtractor ones
and derive reasonable zeropoints for each image. We typically rejected
a few outliers in {\sc fitparams} to obtain an adequate fit (0.3--0.5
mag \hbox{1$\sigma$} error typically on the zeropoint). The
\ion{He}{2}, [\ion{O}{3}], and [\ion{S}{2}] bands were observed
multiple times, so we combined zeropoint estimates to ensure that
magnitudes for the detected sources were consistent to within
$\approx$0.1--0.2 mags across these images. For bands which sampled
comparable spectral windows, we used {\sc synphot} to convert from one
band to another (for instance, from TAURUS 5020\,\AA/30\,\AA~to {\it HST}
F502N or from the VLT FORS spectrum to all overlapping bands). The
resulting magnitudes are listed in Table~\ref{tab:data_aat}, while
Fig.~\ref{fig:aat_images} shows 12 images taken in 1995 February and
1996 March.

\begin{deluxetable*}{lllrll}
\tabletypesize{\scriptsize}
\tablewidth{0pt}
\tablecaption{ESO Observations\label{tab:data_eso}}
\tablehead{
\colhead{Instrument} & 
\colhead{Filter} & 
\colhead{Date} & 
\colhead{Exp.} &
\colhead{Seeing} &
\colhead{Magnitude}}
\tableheadfrac{0.05}
\startdata
SUSI/NTT           & ESO\#369 ([\ion{O}{3}])              & 1993-04-09 & 900 & 0\farcs7 & $>21.7$      \\
SUSI/NTT           & ESO\#430 (5108\,\AA~cont.)               & 1993-04-09 & 900 & 0\farcs7 & ---          \\
SUSI/NTT           & ESO\#629 (H$\alpha$+[\ion{N}{2}])     & 1993-04-09 & 480 & 0\farcs7 & $19.1\pm0.2^{\dagger}$ \\
SUSI/NTT           & ESO\#700 ([\ion{S}{2}])               & 1993-04-09 & 900 & 0\farcs7 & $21.2\pm0.3^{\dagger}$ \\
SUSI/NTT           & ESO\#443 (7027\,\AA~cont.)               & 1993-04-09 & 900 & 0\farcs7 & ---          \\
SUSI/NTT           & ESO\#415 ([\ion{Fe}{11}]$\lambda$7892) & 1993-04-09 & 900 & 0\farcs7 & ---          \\
IRAC2/ESO-MPI 2.2m & J (1.25~$\mu$m)                        & 1993-06-25 & 360 & 0\farcs9 & $>22.0$          \\
IRAC2/ESO-MPI 2.2m & H (1.65~$\mu$m)                        & 1993-06-25 & 360 & 0\farcs9 & $>19.8$          \\
\enddata
\tablecomments{
{\it Column 1:} Instrument and telescope.
{\it Column 2:} Filter.
{\it Column 3:} UT date of observation given as year-month-day.
{\it Column 4:} Exposure time in seconds.
{\it Column 5:} Seeing.
{\it Column 6:} Aperture-corrected Vega magnitude or \hbox{3$\sigma$}
upper limit. Details are given in $\S$~\ref{data_eso}. The detected
magnitudes denoted by $\dagger$s provide constraints on line emission
from the H\,{\sc ii} region which spatially overlaps \hbox{SN\,1996cr}.
}
\end{deluxetable*}

\begin{deluxetable}{lllrl}
\tabletypesize{\scriptsize}
\tablewidth{0pt}
\tablecaption{{\it HST} Observations \label{tab:data_hst}} 
\tablehead{
\colhead{Instrument} & 
\colhead{Filter} & 
\colhead{Date} & 
\colhead{Exp.} & 
\colhead{Magnitude}}
\tableheadfrac{0.05}
\startdata
WFPC2 & F606W & 1996-08-11 & 200, 400 &  --- \\
NIC3  & F164N & 1998-03-16 &       64 & $20.75\pm0.28$ \\
NIC3  & F166N & 1998-03-16 &       64 & $>21.09$ \\
NIC3  & F196N & 1998-03-16 &       80 & $21.04\pm0.29$ \\
NIC3  & F200N & 1998-03-16 &       80 & $20.59\pm0.27$ \\
NIC3  & F212N & 1998-10-16 &      160 & $20.90\pm0.18$ \\
NIC3  & F215N & 1998-10-16 &      160 & $20.60\pm0.19$ \\
WFPC2 & F502N & 1999-04-10 & 900, 900 & $20.67\pm0.06$ \\
WFPC2 & F547M & 1999-04-10 &       60 & $22.81\pm0.12$ \\
WFPC2 & F656N & 1999-04-10 & 800, 800 & $17.98\pm0.11$ \\
WFPC2 & F814W & 1999-04-10 &       40 & $21.20\pm0.09$ \\
\enddata
\tablecomments{
{\it Column 1:} {\it HST} instrument and filter. Filter denotes the
central filter wavelength in nanometers and the filter width
(N=narrow, M=medium, W=wide). Details on individual filters can be
found in the WFPC2\footnote{
\url{http://www.stsci.edu/instruments/wfpc2/Wfpc2\_hand\_current/}}
and NICMOS\footnote{\url{http://www.stsci.edu/hst/nicmos/documents/handbooks/v4.1/}}
Instrument Handbooks. 
{\it Column 2:} UT date of observation given as year-month-day.
{\it Column 3:} Exposure time(s) in seconds.
{\it Column 4:} Aperture-corrected Vega magnitude  or \hbox{3$\sigma$} upper
limit. Magnitudes were measured with circular
apertures of radius 0\farcs2 and 0\farcs5 for WFPC2 and NIC3,
respectively, and corrected for missing flux due to the shape of the
PSF. Note that increasing the aperture radius in the {\it HST} F656N
band to 1\farcs2 yields a magnitude of 17.5 (i.e., an increase of
$\approx$25\% over the point-like magnitude from
\hbox{SN\,1996cr} alone) which we attribute to the flux of the
underlying H\,{\sc ii} region. We have added a systematic error
of 0.1 magnitude to the F656N value to reflect the accuracy to which
we can deblend the point source emission from the underlying H\,{\sc
ii} region.  The H\,{\sc ii} region alone should have a magnitude of
$18.7\pm0.2$, which is consistent with values measured from our
early-time comparison images. Likewise, increasing the aperture in the
NIC3 observations any further will begin to include nearby
contaminating point sources, as shown in
Fig.~\ref{fig:hst_images}. Finally, we have incorporated an additional
systematic error of 0.1 mag to account for the intrapixel sensitivity
variations in the NIC3 observations.}
\end{deluxetable}

\begin{deluxetable}{llrll}
\tabletypesize{\scriptsize}
\tablewidth{0pt}
\tablecaption{AAT TAURUS Observations\label{tab:data_aat}}
\tablehead{
\colhead{Filter} & 
\colhead{Date} & 
\colhead{Exp.} &
\colhead{Seeing} & 
\colhead{Magnitude}}
\tableheadfrac{0.05}
\startdata
5020/30~($\theta_{f}=$0) & 1995-02-21 &  300 & 1\farcs2 & $>20.0$ \\
6583/45~($\theta_{f}=$9) & 1995-02-22 &  300 & 1\farcs2 & $>18.3$ \\
6583/45~($\theta_{f}=$9) & 1995-02-28 &  267 & 1\farcs1 & $>18.1$ \\
6618/55~($\theta_{f}=$7) & 1995-02-28 & 1442 & 1\farcs1 & $>18.4$ \\
$R$                      & 1995-02-28 &   60 & 1\farcs2 & $>19.6$ \\
$B$                      & 1995-02-28 &   20 & 1\farcs4 & $>22.1$ \\
4690/26~($\theta_{f}=$5) & 1996-03-16 & 1200 & 3\farcs5 & $19.5\pm0.6$ \\
6730/30~($\theta_{f}=$5) & 1996-03-16 & 1200 & 3\farcs0 & $15.8\pm0.5$ \\
4690/26~Z-CUBE ($\theta_{f}=$5) & 1996-03-16 & & 3\farcs0 & \\
4690/26~($\theta_{f}=$0) & 1996-03-17 & 1200 & 3\farcs1 & $19.8\pm0.5$ \\
4690/26~($\theta_{f}=$0) & 1996-03-19 & 1000 & 1\farcs2 & $19.7\pm0.3$ \\
5020/30~($\theta_{f}=$0) & 1996-03-19 &  120 & 1\farcs3 & $18.7\pm0.4$ \\ 
6730/30~($\theta_{f}=$0) & 1996-03-19 & 1000 & 1\farcs3 & $15.4\pm0.3$ \\
\enddata
\tablecomments{
{\it Column 1:} Filter, given here as central wavelength and bandwidth
in Angstroms. 6370/30, 6583/45, 5020/30, 4690/26~are narrow-band
filters used with the AAT TAURUS Tunable Filter (see AAT Filter
Catalog\footnote{\url{http://www.aao.gov.au/local/www/cgt/ccdimguide/filtercat.html}}
for details), while $B$ and $R$ are standard Johnson filters. Note
that for the {\it TAURUS} instrument, we also list a tilt angle, which
tunes the wavelength range such that $\lambda_{eff}\approx
\lambda_{nom} (1-\theta_{f}^{2}/28984)$, where $\theta_{f}$ is tilt
angle in degrees.
{\it Column 2:} UT date of observation given as year-month-day.
{\it Column 3:} Exposure time in seconds.
{\it Column 4:} Seeing.
{\it Column 5:} Aperture-corrected Vega magnitude or \hbox{3$\sigma$} upper
limit. Details are given in $\S$~\ref{data_aat}.
}
\end{deluxetable}

\subsubsection{UKST}\label{data_ukst}

The Circinus Galaxy was observed on numerous occasions with the UKST
as detailed in Table~\ref{tab:data_ukst}. The data were retrieved from
the SUPERCOSMOS archive,\footnote{\url{http://surveys.roe.ac.uk/ssa/}}
although only a subset of the photographic plates have been properly
digitized and archived in a usable form. Fig.~\ref{fig:ukst_images}
shows four images taken in 1998 March and April. The H$\alpha$ images
from 1998 (as well as similar degraded ones from 1999 and 2001) show
a slight enhancement over the 1995 H$\alpha$ reference images from the
AAT and ESO. For the four images that were taken after the discovery
window of \hbox{SN\,1996cr} and properly digitized, we performed
photometry using a 1\farcs0 radius circular aperture and previously
established zero points. An aperture correction of 0.2 mags was
estimated empirically using several bright, isolated point sources in
the images. The resulting aperture-corrected magnitudes are presented
in Table~\ref{tab:data_ukst}.

\begin{deluxetable}{lllrll}
\tabletypesize{\scriptsize}
\tablewidth{0pt}
\tablecaption{UKST Observations\label{tab:data_ukst}}
\tablehead{
\colhead{Plate \#} & 
\colhead{Filter} &
\colhead{Date} &
\colhead{Exp.} &
\colhead{Grade} &
\colhead{Magnitude}}
\tableheadfrac{0.05}
\startdata
OR14405 & IIIaF/OG590 & 1991-07-17 &  2400 & AI3* & --- \\
OR16161 & IIIaF/OG590 & 1994-06-27 &  2480 & BI3* & --- \\
OR16238 & IIIaF/OG590 & 1994-08-12 &  3300 & AID2 & --- \\
OR17455 & IIIaF/OG590 & 1997-03-11 &   300 & aI*  & --- \\
OR17484 & IIIaF/OG590 & 1997-03-31 &   300 & a*   & --- \\
HA17930 & 4415/HA659 & 1998-03-05 & 10800 & A2   & $17.1\pm0.6$ \\
OR17954 & 4415/OG590 & 1998-03-19 &   900 & a    & $18.3\pm0.9$ \\
HA17987 & 4415/HA659 & 1998-04-05 & 10800 & AT2  & $17.3\pm0.4$ \\
OR17988 & 4415/OG590 & 1998-04-05 &   900 & a    & $18.0\pm0.7$ \\
OR18316 & 4415/OG590 & 1999-03-11 &   900 & a*   & --- \\
HA18323 & 4415/HA659 & 1999-03-12 &  2160 & bU*  & --- \\
HA18328 & 4415/HA659 & 1999-03-14 & 10800 & bT*  & --- \\
HA19197 & 4415/HA659 & 2001-05-24 &  8400 & AU2* & --- \\
\enddata
\tablecomments{
{\it Column 1:} A one- or two-character code for the filter/emulsion
combination along with a running number for all UKST plates. 
{\it Column 2:} Emulsion and filter. OG590 is red, while HA659 is H$\alpha$.
{\it Column 3:} UT date of observation given as year-month-day.
{\it Column 4:} Exposure time in seconds.
{\it Column 5:} All plates are quality controlled and assigned a
grade. The first letter of the grade (usually A,B,C) indicates the
overall quality of the plate. The subsequent letters indicate specific
defects as follows: I - Denotes image size larger than 40 microns.  T
- Denotes detectable image elongation on most images.  U - Denotes
underexposure (relative to exposure time).  D - Denotes overexposed
plate (high central density). Survey plates are normally also given a
grade number; the lower the number the better the plate quality. An
'A' grade plate scores 3 or lower. Plates denoted by '*' were only
available as lower quality ``Finder'' (GIF) format and thus are not as
photometrically reliable. Additionally, the spatial resolution of
plate OR16238 was too poor to provide a useful comparison magnitude
due to source blending.
{\it Column 7:} Aperture-corrected Vega magnitude. Details are given in
$\S$~\ref{data_ukst}.
}
\end{deluxetable}

\subsubsection{{\it Swift} UVOT}\label{data_swiftuvot}

The Circinus Galaxy was observed on 2007-03-23 with the {\it Swift}
Ultraviolet/Optical Telescope (UVOT) for \hbox{617--2461~s} depending
on the filter. The images were reduced using standard pipeline
procedures and photometry was performed on the images using a 3\farcs0
radius circular aperture. \hbox{SN\,1996cr} was not detected in any of
the six UVOT filters, with aperture-corrected \hbox{3$\sigma$} upper
limits of $V>19.0$, $B>20.2$, $U>20.0$, $UVW1>20.4$, $UVW2>20.8$, and
$UVM2>21.2$.

\subsubsection{SNe Optical Monitoring}\label{data_optmon}

There are three additional useful upper-limit constraints based on the
SNe monitoring observations of the Circinus Galaxy by Rev. Robert
Evans on 1995-03-31, 1995-08-13, and 1996-03-03 (private
communication, 2000). We have adopted upper limits of $V>14$ based on
R. Evans' estimated limiting magnitude of $V\approx15$.

\begin{deluxetable*}{lrrrrrrl}
\tabletypesize{\scriptsize}
\tablewidth{0pt}
\tablecaption{X-ray Observations\label{tab:data_xray}}
\tablehead{
\colhead{Instrument} & 
\colhead{Date} & 
\colhead{Exp.} & 
\colhead{$F_{\rm 0.5-2~keV}$} & 
\colhead{$F_{\rm 2-8~keV}$} &
\colhead{$F^{c}_{\rm 0.5-2~keV}$} & 
\colhead{$F^{c}_{\rm 2-8~keV}$} &
\colhead{Comment}}
\tableheadfrac{0.05}
\startdata
{\it ASCA}           & 1995-02-14 & 61.0 & $<5^{\dagger}$  & $<142^{\dagger}$ & $<19^{\dagger}$  & $<152^{\dagger}$ & \\
{\it ROSAT} HRI      & 1995-09-14 &  4.1 & $<6$            & ---              & $<23$            & ---              & \\ 
{\it ROSAT} HRI      & 1996-02-18 &  1.1 & $<23$           & ---              & $<93$            & ---              & \\ 
{\it ROSAT} HRI      & 1996-09-13 &  1.8 & $<11$           & ---              & $<44$            & ---              & \\ 
{\it ROSAT} HRI      & 1997-03-03 & 26.4 & $<2$            & ---              & $<8$             & ---              & \\ 
{\it ROSAT} HRI      & 1997-08-17 & 45.9 & $<1$            & ---              & $<5$             & ---              & \\ 
{\it BeppoSAX}       & 1998-03-13 & 85.2 & $<14^{\dagger}$ & $<154^{\dagger}$ & $<51^{\dagger}$  & $<165^{\dagger}$ & \\
{\it Chandra} ACIS-S & 2000-01-16 &  1.0 & $13\pm3$        & $142\pm35$       & $62\pm16$        & $142\pm35$       & 21\% pile-up\\ 
{\it Chandra} ACIS-S & 2000-03-14 &  4.9 & $14\pm3$        & $133\pm15$       & $51\pm10$        & $140\pm16$       & 2\% pile-up\\ 
{\it Chandra} ACIS-S & 2000-03-14 & 23.1 & $10\pm2$        & $120\pm20$       & $45\pm9$         & $120\pm20$       & 20\% pile-up \\ 
{\it Chandra} HETGS  & 2000-06-15 & 67.1 & $12\pm2$        & $108\pm9$        & $43\pm9$         & $118\pm9$        & 3\% pile-up\\ 
{\it BeppoSAX}       & 2001-01-07 & 27.2 & $<80^{\dagger}$ & $<614^{\dagger}$ & $<292^{\dagger}$ & $<620^{\dagger}$ & \\
{\it Chandra} ACIS-S & 2001-05-02 &  4.4 & $12\pm2$        & $136\pm20$       & $53\pm11$        & $136\pm20$       & 21\% pile-up \\ 
{\it XMM-Newton}     & 2001-08-06 & 85.5 & $16\pm1$        & $144\pm5$        & $61\pm2$         & $154\pm5$        & \\
{\it Chandra} HETGS  & 2004-06-02 & 55.0 & $23\pm3$        & $186\pm11$       & $88\pm12$        & $204\pm11$       & 3\% pile-up \\ 
{\it Chandra} HETGS  & 2004-11-28 & 59.0 & $24\pm3$        & $196\pm12$       & $93\pm11$        & $215\pm13$       & 3\% pile-up \\ 
{\it Swift} XRT      & 2007-03-25 &  8.0 & $35\pm8$        & $206\pm39$       & $124\pm28$       & $222\pm42$       & \\
\enddata
\tablecomments{
{\it Column 1:} Satellite and instrument. 
{\it Column 2:} Starting date of observation.
{\it Column 3:} Exposure time in ksec.
{\it Columns 4 and 5:} Flux in the 0.5--2~keV and 2--8~keV bands
respectively, determined from the best-fit model in {\sc xspec} in
units of $10^{-14}$~erg~s$^{-1}$~cm$^{-2}$. For upper limits, we have
adopted the spectrum from the closest {\it Chandra} observation. For
instruments unable to resolve \hbox{SN\,1996cr} from the bright nucleus
of the Circinus Galaxy, denoted by $^{\dagger}$, upper limit fluxes
were estimated by subtracting the total flux of the Circinus Galaxy
without \hbox{SN\,1996cr} from the total unresolved flux (see
$\S$\ref{data_xray} for details and caveats).
{\it Columns 6 and 7:} Absorption-corrected flux in the 0.5--2~keV
and 2--8~keV bands respectively, determined from the best-fit model
in {\sc xspec} in units of $10^{-14}$~erg~s$^{-1}$~cm$^{-2}$. 
{\it Column 8:} Comments.
}
\end{deluxetable*}

\subsection{Archival X-ray Data}\label{data_xray}

Details of the various \hbox{X-ray} observations are given in
Table~\ref{tab:data_xray}. We describe below our reduction methods for
each dataset.  When possible, the \hbox{X-ray} fluxes and
absorption-corrected luminosities for \hbox{SN\,1996cr} were calculated via
spectral analysis in {\sc xspec} \citep{Arnaud1996} using the Cash
statistic \citep{Cash1979}. Unless stated otherwise, errors on
spectral parameters are for 68\% confidence, assuming one parameter of
interest. Following the treatment of \hbox{SN\,1987A}
\citep[e.g.,][]{Park2005}, we characterized the \hbox{X-ray} spectra of
\hbox{SN\,1996cr} with an absorbed variable-abundance non-equilibrium
ionization (NEI) shock model ({\it vpshock} using NEI version 2.0 in
{\sc xspec}) based on ATOMDB \citep[][]{Smith2001}.\footnote{While
this model is believed to best characterize the \hbox{X-ray} emission from
young SNe such as \hbox{SN\,1996cr}, there are still many serious caveats. See
\url{http://cxc.harvard.edu/atomdb/issues\_caveats.html} for details.}
The best-fitted parameters to the combined {\it XMM-Newton} and {\it
Chandra} HETGS 0th order dataset were as follows: $N_{\rm
H}=8.4\times10^{21}$~cm$^{-2}$, $kT=13.4$~keV, $Z_{\rm Si}=2.25 Z_{\rm
\odot,Si}$, $Z_{\rm S}=3.5 Z_{\rm \odot,S}$, $Z_{\rm Ca}=8 Z_{\rm
\odot,Ca}$, and $Z_{\rm Fe}=2 Z_{\rm \odot,Fe}$. We note, however, 
that these specific abundances, and even the estimated temperature to
some extent, should be used with caution, as there are known to be
significant deviations between adopted
models \citep[e.g.,][]{Nymark2006}. The typical column density,
$N_{\rm H}$, derived from spectral fits to the data is much larger
than the estimated Galactic absorption column
($\approx$3$\times10^{21}$~cm$^{-2}$), implying significant internal
absorption either from the disk of the Circinus Galaxy at large or the
immediate vicinity of
\hbox{SN\,1996cr}. Fluxes were measured by varying the normalization of
the above model in the \hbox{0.5--2~keV} or \hbox{2--8~keV} bands. Our
adopted model fit all of the spectral data adequately, although we
note that there was typically some excess residual emission around
prominent emission lines, as well as an apparent
\hbox{$\sim2\sigma$} soft excess in the 2004 HETGS data. These
issues have a negligible effect, however, on our flux estimates to
within errors and will be addressed in more detail in a separate
publication on \hbox{X-ray} spectral analysis of \hbox{SN\,1996cr}
(F. Bauer et al., in preparation).

\subsubsection{{\it Chandra}}\label{data_chandra}

The Circinus Galaxy was observed on several occasions with {\it
Chandra} using the backside-illuminated Advanced Imaging CCD
Spectrometer (ACIS-S) in the focal plane, both with and without the
High-Energy Transmission Grating Spectrometer (HETGS). Archival ACIS-S
and HETGS zeroth-order data were retrieved from the {\it Chandra} Data
Archive and processed following standard procedures using {\sc ciao}
(v3.4) software. Additionally we removed the 0\farcs5 pixel
randomization, corrected for charge transfer inefficiency (CTI),
performed standard {\it ASCA} grade selection, excluded bad pixels and
columns, and screened for intervals of excessively high background
(none was found).  Analysis was performed on reprocessed {\it
Chandra} data, primarily using {\sc ciao}, but also with {\sc ftools}
(v6.3) and custom software including {\sc acis extract}
\citep[v3.128;][]{Broos2007}. Spectra were extracted using {\sc acis
extract} with a 95\% encircled-energy region derived from the {\it
Chandra} PSF library. As \hbox{SN\,1996cr} lies at the edge of diffuse
emission associated with the AGN, a background spectrum was extracted
from a local annular region after excluding nearby point sources and
the strongest portions of the circumnuclear halo and ionization cone
associated with the AGN (the annulus size was grown until
approximately 100 counts could be extracted). We generated calibration
products (including an aperture correction) for the HETGS/ACIS-S
spectra and fit them separately within {\sc xspec} using our best-fit
model to estimate the flux. The majority of the observations were
performed in configurations which mitigate pile-up; when significant
we have noted the estimated pile-up fractions for
\hbox{SN\,1996cr}, and corrected for them using the {\sc xspec} {\it pileup}
model of \citet{Davis2001}.

\subsubsection{{\it XMM-Newton}}\label{data_xmm}

The Circinus Galaxy was observed once with {\it XMM-Newton} using the
EPIC p-n and MOS1/MOS2 detectors on 2001-08-06 for
$\approx104$~ks. Archival data were retrieved from the {\it XMM-Newton}
Science Archive and processed following standard procedures with SAS
(v7.0.0). Additionally, we screened for and removed intervals of
excessively high background ($\approx15$\% for MOS1/MOS2 and
$\approx40$\% for p-n), leaving 85.5~ks, 91.8~ks, and 59.5~ks of
useful exposure with the MOS1, MOS2, and p-n instruments,
respectively. Spectra were extracted with SAS using a circular
aperture of radius 11\farcs0 and local background was extracted in a manner
similar to that done for the {\it Chandra} observations. We generated
calibration products for the p-n/MOS1/MOS2 datasets and joint-fit the
three spectra within {\sc xspec} using our best-fit model to estimate
the flux.

Additionally, we used the {\it XMM-Newton} data to place upper limits
on the potential flux from \hbox{SN\,1996cr} in the {\it ASCA} and {\it
BeppoSAX} datasets, both of which lack the spatial resolution to
resolve the galaxy into its various components \citep[e.g., see
Fig.~1 of][]{Bauer2001}. To this end, spectra for the entire
Circinus Galaxy were extracted from a 2\farcm5 radius circular
aperture, excluding emission from \hbox{SN\,1996cr}. Unfortunately, the
ultraluminous \hbox{X-ray} source \hbox{CG~X-1} is known to vary with a
period of 7.5~hr and comprises a non-negligible fraction of the total
\hbox{X-ray} emission from Circinus. Thus we extracted a pessimistic
spectrum additionally excluding \hbox{CG~X-1}. This provides conservative
upper limits for the total emission from the Circinus Galaxy during
the {\it ASCA} and {\it BeppoSAX} observations, under the assumption
that the fluxes of the various emission components within the Circinus
Galaxy (aside from \hbox{CG~X-1} and \hbox{SN\,1996cr}) have not changed since
1995. Based on the spatially resolved observations that we do have,
this assumption appears reasonable. More optimistic upper limits can
be estimated by further subtracting the time-averaged flux from
\hbox{CG~X-1}. Given ambiguities in determining how the unresolved
observations sample the light curve of \hbox{CG~X-1}, we simply
provide here the flux of \hbox{CG~X-1} in its ``high'' state
\citep[see ][]{Bauer2001} which is
$5.7\times10^{-13}$~ergs~s$^{-1}$~cm$^{-2}$ and
$4.3\times10^{-12}$~ergs~s$^{-1}$~cm$^{-2}$ in the 0.5--2~keV and
2--8~keV bands, respectively, and note that optimistic upper limits
for the flux of \hbox{SN\,1996cr} would likely be a factor of
$\approx$2--3 lower.

\subsubsection{{\it ROSAT}}\label{data_rosat}

The Circinus Galaxy was observed on several occasions with {\it ROSAT}
using the High-Resolution Imager (HRI). Processed data were retrieved
from the High Energy Astrophysics Science Archive Research Center
(HEASARC), and analysis was performed using {\sc ftools} and custom
software. We adopted the best-fit model from the combined observations
(see above) to estimate the {\it ROSAT} HRI upper limits. Upper limits
were measured using a 10$\arcsec$ aperture and a local background
extracted to match roughly the amount of diffuse and scattered
emission thought to reside within the source aperture. Count rates
were converted to fluxes with {\sc xspec} using the final {\it ROSAT}
calibration products and our adopted best-fit model. Adopting similar
models does not have a significant impact on our derived flux upper
limits.

\subsubsection{{\it Swift}}\label{data_swift}

The Circinus Galaxy was observed once with the {\it Swift}
\hbox{X-ray} Telescope (XRT) on 2007 March 23 for 7.4~ks. Processed
data were retrieved from {\it Swift} Archive, and analysis was
performed using {\sc ftools} and custom software. \hbox{SN\,1996cr} is
visibly separated from the nucleus (25\arcsec separation) and diffuse
circumnuclear emission and appears to be only marginally
contaminated. A spectrum of \hbox{SN\,1996cr} was extracted from the
event list using a 10~pixel (23\farcs5) radius circular aperture,
masked to exclude emission from the nucleus and diffuse halo to as
large an extent as possible. Additionally, a background was extracted
in a manner similar to that done for the {\it Chandra} observations so
as to match roughly the amount of diffuse and scattered emission
believed to be present within the source aperture. Low photon
statistics, however, limited the fidelity of the background
subtraction. We generated calibration products for the spectrum and
fit it within {\sc xspec} using our best-fit model to estimate the
flux.

\subsubsection{{\it ASCA}}\label{data_asca}

The Circinus Galaxy was observed once with {\it ASCA} using the Gas
Imaging Spectrometers (GIS) and Solid State Imaging Spectrometers (SIS)
on 1995-02-14 for 61.1~ks. Processed data were retrieved from
HEASARC, providing \hbox{$\approx$23--36~ks} of usable data.  Analysis was
performed using {\sc ftools} and custom software. The target appears
unresolved with {\it ASCA} and was treated as a point source for
extraction purposes. Spectra for the entire Circinus Galaxy were
extracted from a 2\farcm5 radius circular aperture, while backgrounds
were taken from blank-sky observations at the same position and with
identical screening criteria. Contamination from a neighboring source
5$\arcmin$ away was minimal. Total fluxes were measured via
simultaneous spectral fitting of the GIS and SIS data and upper limits
for \hbox{SN\,1996cr} were estimated by subtracting the flux estimated
from the {\it XMM-Newton} data as detailed above.

\subsubsection{{\it BeppoSAX}}\label{data_sax}

The Circinus Galaxy was observed on two occasions with {\it BeppoSAX}
using the Low Energy Concentrator Spectrometer (LECS) and Medium
Energy Concentrator Spectrometer (MECS) on 1998-03-03 for
83.9/138.0~ks and on 2001-01-07 for 26.9/52.1~ks. Processed data
were retrieved from HEASARC, providing 83.7/137.2~ks and 26.8/51.7~ks
of usable data, respectively. Analysis was performed using {\sc
ftools} and custom software. The target appears unresolved with {\it
BeppoSAX} and was treated as a point source for extraction
purposes. LECS and MECS spectra for the entire Circinus Galaxy were
extracted from a 3--4$\arcmin$ radius circular apertures (limited by
the availability of calibration files), while backgrounds were taken
from blank-sky observations at the same position and with identical
screening criteria. Contamination from a neighboring source 5$\arcmin$
away was minimal. Total fluxes were measured via simultaneous spectral
fitting of the LECS and MECS data, and upper limits for
\hbox{SN\,1996cr} were estimated by subtracting the flux estimated from
the {\it XMM-Newton} data as detailed above.

\subsection{Archival and Proposed Radio Data}\label{data_radio}

\begin{deluxetable*}{lrr|rr|rr|rr|rr|rr|l}
\tabletypesize{\scriptsize}
\tablewidth{0pt}
\tablecaption{ATCA Radio Observations\label{tab:data_radio}}
\tablehead{
\colhead{Obs. Date} &
\colhead{Obs. ID} &
\colhead{Array} &
\multicolumn{2}{|c|}{K-Band} &
\multicolumn{2}{|c|}{X-Band} &
\multicolumn{2}{|c|}{C-Band} &
\multicolumn{2}{|c|}{S-Band} &
\multicolumn{2}{|c|}{L-Band} &
\colhead{Comments}\\
&
&
&
\colhead{$\nu$} &
\colhead{$S_{\nu}$} &
\colhead{$\nu$} &
\colhead{$S_{\nu}$} &
\colhead{$\nu$} &
\colhead{$S_{\nu}$} &
\colhead{$\nu$} &
\colhead{$S_{\nu}$} &
\colhead{$\nu$} &
\colhead{$S_{\nu}$} &
}
\tableheadfrac{0.05}
\startdata
1995-03-03 & C204  & 375   &   --- &        --- &  8640 &      $>$0.7 &  4800 &      $>$1.3 &   --- &       --- &   --- &        --- & mosaic field ``a'' \\
1995-03-03 & C204  & 375   &   --- &        --- &  8640 &      $>$0.7 &  4800 &      $>$0.9 &   --- &       --- &   --- &        --- & mosaic field ``b'' \\
1995-03-30 & C363  & 1.5A  &   --- &        --- &   --- &         --- &   --- &         --- &   --- &       --- &  1418 &     $>$6.6 & \\
1995-04-16 & C418  & 6C    &   --- &        --- &   --- &         --- &   --- &         --- &  2768 &    $>$1.1 &  1418 &     $>$1.2 & \\
1995-05-27 & C204  & 375   &   --- &        --- &  8640 &      $>$0.7 &  4800 &      $>$1.3 &   --- &       --- &   --- &        --- & mosaic field ``a'' \\
1995-05-27 & C204  & 375   &   --- &        --- &  8640 &      $>$0.7 &  4800 &      $>$0.6 &   --- &       --- &   --- &        --- & mosaic field ``b'' \\
1995-06-07 & C204  & 375   &   --- &        --- &  8640 &      $>$0.7 &  4800 &      $>$1.5 &   --- &       --- &   --- &        --- & mosaic field ``a'' \\
1995-06-07 & C204  & 375   &   --- &        --- &  8640 &      $>$1.5 &  4800 &      $>$0.5 &   --- &       --- &   --- &        --- & mosaic field ``b'' \\
1995-07-27 & C204  & 6C    &   --- &        --- &   --- &         --- &   --- &         --- &  2368 &    $>$1.3 &  1376 &     $>$3.4 & \\
1995-08-01 & CT09  & 750B  &   --- &        --- &   --- &         --- &   --- &         --- &   --- &       --- &  1418 &     $>$17  & \\
1996-02-06 & C466  & 750D  &   --- &        --- &   --- &         --- &   --- &         --- &   --- &       --- &  1664 &     $>$1.5 & \\
1996-08-01 & C363  & 375   &   --- &        --- &  8640 & 0.5$\pm$0.1 &  4800 &      $>$1.2 &   --- &       --- &   --- &        --- & \\
1996-12-15 & C586  & 375   &   --- &        --- &  8512 & 2.2$\pm$0.3 &  4928 & 1.8$\pm$0.2 &   --- &       --- &   --- &        --- & mosaic field ``a'' \\
1996-12-15 & C586  & 375   &   --- &        --- &  8512 & 2.0$\pm$0.2 &  4928 & 1.8$\pm$0.2 &   --- &       --- &   --- &        --- & mosaic field ``b'' \\
1996-12-05 & C505  & 375   &   --- &        --- &   --- &         --- &   --- &         --- &  2368 &    $>$0.4 &  1384 &     $>$0.7 & \\
1997-06-17 & V100  & 6A    &   --- &        --- &  8425 &  14$\pm$1   &   --- &         --- &   --- &       --- &   --- &        --- & \\
1997-06-17 & V100  & 6A    &   --- &        --- &  8425 &  15$\pm$3   &   --- &         --- &   --- &       --- &   --- &        --- & spectral-line mode \\
1997-12-05 & V099  & 6C    &   --- &        --- &   --- &         --- &   --- &         --- &  2268 &  22$\pm$3 &   --- &        --- & \\
1997-12-05 & V099  & 6C    &   --- &        --- &   --- &         --- &   --- &         --- &  2268 &  37$\pm$4 &   --- &        --- & spectral-line mode \\
1997-12-31 & C694  & 6C    &   --- &        --- &  8500 &  96$\pm$15  &  4800 & 119$\pm$12  &   --- &       --- &   --- &        --- & 8.5~GHz extrapolated \\
           &       &       &       &            &       &             &       &             &       &           &       &            & from 6.0~GHz data\\
1999-06-19 & C788  & 375   &   --- &        --- &   --- &         --- &   --- &         --- &   --- &       --- &  1384 &   89$\pm$3 & \\
1999-06-19 & C788  & 375   &   --- &        --- &   --- &         --- &   --- &         --- &   --- &       --- &  1418 &   85$\pm$3 & spectral-line mode \\
2000-06-19 & V137  & 6B    & 22190 &  76$\pm$21 &   --- &         --- &   --- &         --- &   --- &       --- &   --- &        --- & \\ 
2000-06-19 & V137  & 6B    & 22203 &  72$\pm$20 &   --- &         --- &   --- &         --- &   --- &       --- &   --- &        --- & \\ 
2000-06-19 & V137  & 6B    & 22215 &  69$\pm$21 &   --- &         --- &   --- &         --- &   --- &       --- &   --- &        --- & \\ 
2002-07-18 & V137  & 1.5G  & 22172 &  63$\pm$10 &   --- &         --- &   --- &         --- &   --- &       --- &   --- &        --- & \\
2002-07-18 & V137  & 1.5G  & 22216 &  71$\pm$10 &   --- &         --- &   --- &         --- &   --- &       --- &   --- &        --- & \\
2002-07-18 & V137  & 1.5G  & 22224 &  70$\pm$10 &   --- &         --- &   --- &         --- &   --- &       --- &   --- &        --- & \\ 
2003-09-07 & C1224 & EW367 & 23659 &  63$\pm$10  &   --- &         --- &   --- &         --- &   --- &       --- &   --- &        --- & phase cal only, flux forced \\
2003-11-03 & C1049 & H214  & 16960 &  80$\pm$10 &   --- &         --- &   --- &         --- &   --- &       --- &   --- &        --- & \\
2003-11-03 & C1049 & H214  & 19008 &  77$\pm$10 &   --- &         --- &   --- &         --- &   --- &       --- &   --- &        --- & \\
2003-11-04 & C1049 & H214  & 21056 &  79$\pm$10 &   --- &         --- &   --- &         --- &   --- &       --- &   --- &        --- & \\
2003-11-04 & C1049 & H214  & 22796 &  73$\pm$10 &   --- &         --- &   --- &         --- &   --- &       --- &   --- &        --- & \\
2003-11-08 & C1049 & 1.5D  &   --- &        --- &  8256 & 163$\pm$10  &  5056 & 227$\pm$11  &   --- &       --- &   --- &        --- & \\
2004-04-02 & C1424 & 6A    &   --- &        --- &  8640 & 162$\pm$8   &  4800 & 240$\pm$10  &  2368 & 364$\pm$16 &  1384 & 481$\pm$19 & \\
2004-05-20 & CX065 & 1.5B  & 22088 &  63$\pm$15 &   --- &         --- &   --- &         --- &   --- &       --- &   --- &        --- & phase cal only, flux forced \\
2004-06-13 & V176  & 750D  & 22209 &  57$\pm$13 &   --- &         --- &   --- &         --- &   --- &       --- &   --- &        --- & \\
2004-06-13 & V176  & 750D  & 22195 &  59$\pm$11 &   --- &         --- &   --- &         --- &   --- &       --- &   --- &        --- & \\
2004-08-01 & V176B & H168  & 22209 &  66$\pm$6  &   --- &         --- &   --- &         --- &   --- &       --- &   --- &        --- & \\
2004-08-01 & V176B & H168  & 22195 &  66$\pm$6  &   --- &         --- &   --- &         --- &   --- &       --- &   --- &        --- & \\
2005-03-14 & C1368 & H214  & 22193 &  51$\pm$15 &   --- &         --- &   --- &         --- &   --- &       --- &   --- &        --- & \\
2005-04-20 & C1368 & 750A  & 22194 &  68$\pm$6  &   --- &         --- &   --- &         --- &   --- &       --- &   --- &        --- & phase cal only \\
2005-11-13 & C1049 & 1.5C  &   --- &        --- &  8640 & 159$\pm$5   &  4800 & 246$\pm$12  &   --- &       --- &   --- &        --- & \\
2006-04-29 & C1049 & H214  & 18752 &  77$\pm$7  &   --- &         --- &   --- &         --- &   --- &       --- &   --- &        --- & \\
2006-04-29 & C1049 & H214  & 21056 &  67$\pm$6  &   --- &         --- &   --- &         --- &   --- &       --- &   --- &        --- & \\
2006-06-21 & C1049 & 1.5D  &   --- &        --- &  8640 &    154$\pm$11 &  4800 &   261$\pm$10 &   --- &       --- &   --- &        --- & \\
2006-09-14 & C1341 & H75   &   --- &        --- &   --- &         --- &   --- &         --- &   --- &       --- &  1418 &     505$\pm$20 &  \\
2007-06-24 & VX013A & --- & 22316 &  67$\pm$6  &   --- &         --- &   --- &         --- &   --- &       --- &   --- &        --- & \\
\enddata
\tablecomments{
{\it Column 1:} Starting date of observation.
{\it Column 2:} Observing program ID.
{\it Column 3:} Array configuration.
{\it Columns 4, 6, 8, 10, and 12:} Mean frequency of observed band in
units of MHz.
{\it Columns 5, 7, 9, 11, and 13:} Integrated flux densities
(determined from {\sc imfit}) or \hbox{3$\sigma$} upper limits (determined
from {\sc imstat}) in units of mJy. Errors include both statistical and
systematic terms. The systematic error is
estimated from the ratio of the measured calibrator
fluxes over its estimated historical value based on monitored
light curves (typical variance was 5--20\%).
{\it Column 14:} Comments. Observations taken as mosaics, in
spectral-line mode, or with limited calibration sources are duly
noted.
}
\end{deluxetable*}

\subsubsection{Australia Telescope Compact Array}\label{data_atca}

\begin{figure*}
\vspace{-0.1in}
\centerline{
\includegraphics[height=19.0cm]{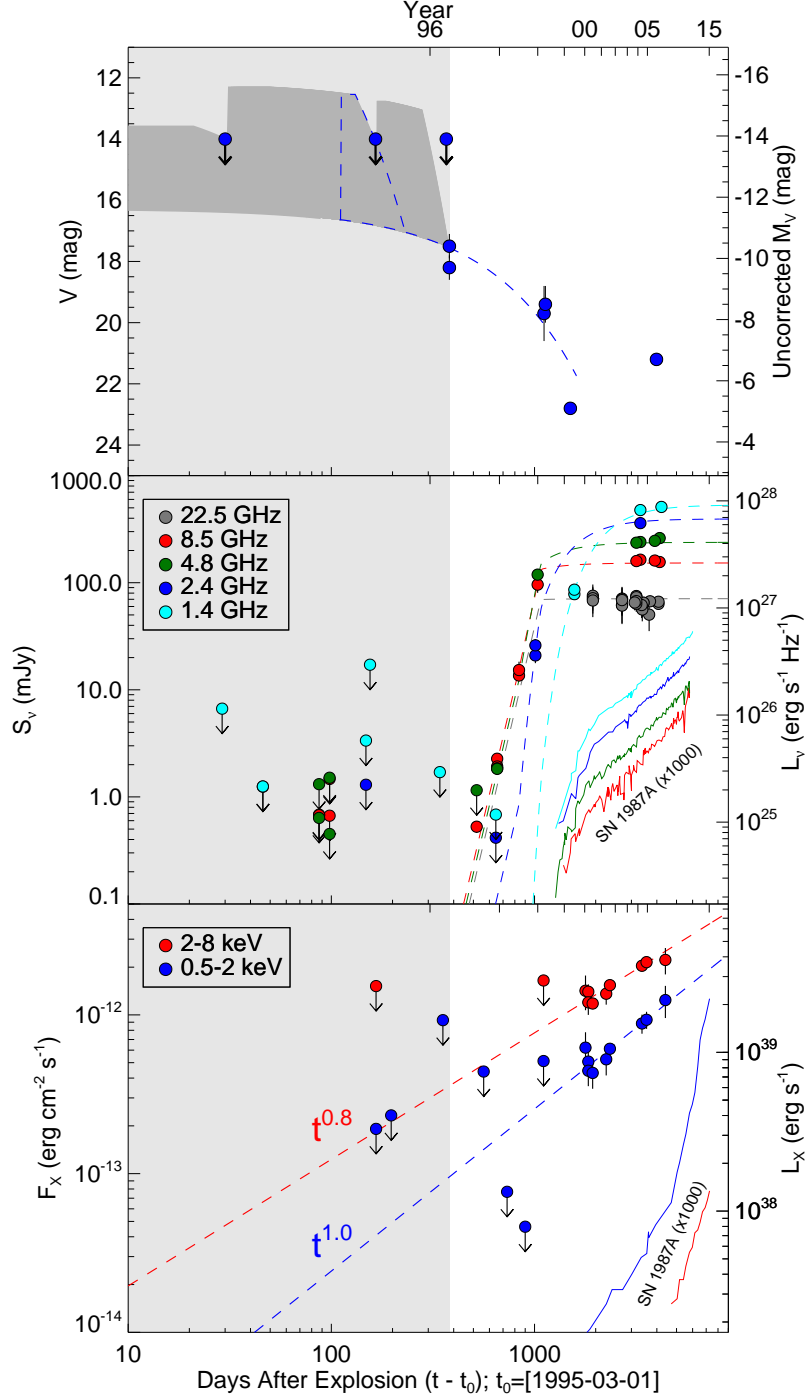}
}
\vspace{-0.0in}
\caption[sn1996cr_lc2.eps]{
%
%
The solid circles denote the light curves for $V$-band (top panel),
radio (middle panel), and absorption-corrected \hbox{X-ray} (bottom
panel) data. Fluxes are shown on the left, while luminosities are on
the right. The light grey shaded region on the left denotes the
current explosion window.
For the $V$-band, we also show a typical $V$-band light curve based on
five \hbox{type~IIn} SNe (dashed blue line). There is a large range in
properties at early times, for which we show only the upper and lower
extremes expected; \hbox{SN\,1996cr}'s maximum light could lie
anywhere in between. We have normalized to fit the initial discovery
magnitudes and adopted a late-time decline rate of $\Delta V=0.0035$
per day to fit the {\it HST} data. None of the magnitudes has been
corrected for extinction, which is likely to be in excess of $A_{\rm
V}\approx4.5$. The dark grey region denotes the range of parameter
space where our adopted SN light curve is not excluded.
The radio data provide relatively robust constraints on emission at
early times throughout the explosion window, and only peak up in
late 1996. The radio emission is fit with a modified ejecta-CSM
interaction model (dashed lines, see $\S$~\ref{radio_ltcrv}), which
deviates from the standard model in two important ways. First, the
unusual sharp rise and variable spectral index cannot be fit by
free-free absorption alone, and requires a $\ga$100 fold increase in
radio luminosity as well. Second, the late-time slope remains flat up
to the current epoch, rather than exhibiting the characteristic
rollover seen in most RSNe.
The late-time \hbox{X-ray} data show a relatively strong and
significant rise between 2000--2007 both in the 0.5--2~keV and
2--8~keV bands. This is very atypical of \hbox{X-ray} detected SNe and
suggests that the blast wave of \hbox{SN\,1996cr} is encountering a
dense circumstellar shell. The dashed lines show empirical power-law
fits to the data with labeled slopes. The early-time upper limits in
all bands suggest that either very strong absorption played a role
early on or more likely the progenitor of \hbox{SN\,1996cr} formed a
cavity during the final part of its evolution. The thin solid curves,
the colors of which mirror those describing the \hbox{SN\,1996cr} data,
show the corresponding X-ray and radio light curves of \hbox{SN\,1987A}
for comparison \citep{Hasinger1996, Manchester2002, Park2006}.  Only
the luminosity and days after explosion axes apply to
the \hbox{SN\,1987A} data. The \hbox{SN\,1987A} data have additionally
been scaled up by a factor of 1,000 for illustrative purposes. While
many of the specifics are different, the general trend of
the \hbox{SN\,1996cr} X-ray and radio light curves resembles those
of \hbox{SN\,1987A}. 
\label{fig:sn1996cr-lc}}
\end{figure*}

Radio observations of the Circinus Galaxy spanning 1995 to 2006 were
retrieved from the Australia Telescope Compact Array (ATCA)
archive\footnote{\url{http://atoa.atnf.csiro.au/}} and are presented
in Table~\ref{tab:data_radio} and Fig.~\ref{fig:sn1996cr-lc}. The data
were reduced with {\sc miriad} (v4.0.5) following the procedures
outlined in the ATNF Miriad User
Manual.\footnote{\url{http://www.atnf.csiro.au/computing/software/miriad/}}

The ``primary'' flux calibrator is 1934$-$638, which is assumed to be
constant in time with flux densities of 14.94, 11.60, 5.83, 2.84, and
1.03~Jy at 20, 13, 6, 3, 1~cm, respectively. As noted in
Table~\ref{tab:data_radio}, there are a few instances in which
1938$-$638 is not included as part of the dataset, although the
observations are initially ``primed'' using this source; for such
cases, fluxes of the ``secondary'' calibrators were compared to their
interpolated historical values from calibration monitoring
efforts\footnote{\url{http://www.narrabri.atnf.csiro.au/calibrators/}}
and were fixed to those values when they strayed by more than
$\sim20\%$ from them. The secondary calibrators, which served as the
gain, bandpass, and phase calibrators for the Circinus Galaxy, varied
from observation to observation and include sources 1921$-$293
(bandpass, alternate 1~cm flux), 1329$-$665 (bandpass), 1549$-$790
(bandpass), 1414$-$59 (phase), and 1236$-$684 (phase). Both source and
calibration data were flagged to remove bad time intervals and
channels, time intervals strongly affected by interference, and
obvious emission and absorption lines. The calibration tables were
applied to the Circinus Galaxy data, from which deconvolved,
primary-beam-corrected images were made. On rare occasions, a phase
self-calibration was required. Integrated flux densities (determined
from {\sc imfit}) or
\hbox{3$\sigma$} upper limits (determined from {\sc imstat}) were then
determined for \hbox{SN\,1996cr}. These values are provided in
Table~\ref{tab:data_radio}.  There appears to be considerable
dispersion in 22.5~GHz flux densities, which were mostly acquired as
part of larger Australia Long Baseline Array observations. We lack
comparison data at other wavelengths to determine if this is simply
additional calibration uncertainty or potentially real flux
variations. However, given that there appears to be dispersion even
between observations taken simultaneously, we caution that the
dispersion is not intrinsic, but rather due to systematic errors.

We note that the only other strong point source in the field of view
was the nucleus of the Circinus Galaxy, which is likely due to
emission from the Compton-thick AGN or a compact circumnuclear
starburst. As a check on the flux of \hbox{SN\,1996cr}, we extracted flux
measurements from the point-like nucleus in a manner identical to
\hbox{SN\,1996cr}. While the nucleus could vary intrinsically and may
additionally be contaminated at long wavelengths by diffuse emission
from the extended disk, it still provides us with a secondary estimate
of any systematic error associated with a given observation. We find
that the majority of measurements of the nucleus lie between 10--30\%
of its mean total flux in each band, although a few observations
(mainly at 22.5~GHz where contamination from water maser emission is
possible) vary by a factor of up to four; we have adopted larger
systematic errors for these observations as a result. Thus, aside from
the 22.5~GHz data, we are generally confident in our measured fluxes
for \hbox{SN\,1996cr}.

\subsubsection{Australia Long Baseline Array}\label{data_lba}

We carried out Very Long Baseline Interferometry (VLBI)
observations of \hbox{SN\,1996cr} on 2007-06-24 using the following
telescopes of the Australian Long Baseline Array: ATCA, Mopra, Parkes,
Hobart, and Ceduna.\footnote{See http://www.atnf.csiro.au/vlbi/ for
details.} The observations were phase-referenced to the nearby
calibrator \hbox{PMN~J1355$-$6326}.  We observed both senses of
circular polarization at a frequency of 22~GHz with a total bandwidth
of 65~MHz, and for a total time of $\sim$9~h. We report here only our
preliminary results and note that a full description is forthcoming
(N. Bartel et al., in preparation).  \hbox{SN\,1996cr} was detected
only on the three baselines involving the Parkes, ATCA and Mopra
telescopes. The flux of \hbox{SN\,1996cr} is consistent with previous
ATCA measurements, while its extent appears resolved by the longest
baselines at $\ga3\sigma$. Although Gaussian and disk models provide
adequate fits, we adopt a model consisting of an optically thin
spherical shell, as was found to be appropriate
for \hbox{SN\,1993J} \citep[see,
e.g.,][]{Bietenholz2001,Bartel2002}. For a ratio of the outer to the
inner shell radius of 1.25, we find an outer angular radius of
$\sim$5~mas, which corresponds to $\sim2.8\times10^{17}$~cm for an
assumed distance of 3.8~Mpc. Modeling a smaller outer to inner shell
ratio like 1.10 could perhaps narrow the angular radius by $\sim$5\%,
while using a filled-center model (equivalent to an inner shell
radius of 0) might increase the angular radius by $\sim$20\%.

\subsection{Archival Gamma-Ray Burst Data}\label{data_grb}

Among the numerous GRB-associated SNe that now exist, nearly all have
been identified as type~Ibc's, which suggests that they arise from
Wolf-Rayet progenitors.  Although SN\,1996cr is identified as a
type~IIn SN at late times, it is worth searching for any evidence of
temporal and spatial gamma-ray emission, as its original type may have
been quite different. Additionally, there are at least two events in
which a good observational case can be made for the association of a
GRB with a type IIn SN:
\hbox{GRB~011121}/\hbox{SN\,2001ke} \citep{Garnavich2003} and
\hbox{GRB~970514}/\hbox{SN\,1997cy} \citep{Germany2000}. 
We find eight BATSE GRB detections over the lifetime of the {\it
Compton} Gamma-Ray Observatory \citep{Meegan1996} which include the
Circinus Galaxy within their \hbox{3$\sigma$} error regions. One of
these GRBs, \hbox{4B~960202}, appears to lie within the explosion date
constraints of \hbox{SN\,1996cr}, suggesting a tentative
association. Importantly, however, \hbox{4B~960202} was detected by
multiple Gamma-ray observatories and has a dramatically smaller error
estimate via triangulation from the Interplanetary
Network \citep[IPN;][]{Hurley1999, Hurley2005}. The full IPN 3$\sigma$
confidence annulus is only $\approx$0\fdg33 wide and lies $\ga$1\fdg3
from the SN at its closest tangential point, thus strongly ruling out
any identification with \hbox{SN\,1996cr}.

\section{Isolation of \hbox{SN\,1996\MakeLowercase{cr}}'s Explosion Date}\label{confirm}

Tables~\ref{tab:data_aat}--\ref{tab:data_radio} give a detailed
summary of the observations relevant for determining the explosion
date of \hbox{SN\,1996cr}. While \hbox{SN\,1996cr} was discovered in the
X-ray band, the archival \hbox{X-ray} data available are too limited
and subject to potential geometrical and structural effects (e.g.,
early absorption, cavities) to narrow down the explosion date to
better than an $\sim10$~yr window between 1990--2000. Although
\hbox{SN\,1996cr} is relatively well-sampled at radio frequencies,
those observations similarly suffer from the same early-time effects,
thus limiting their utility to date the SN. Fig.~\ref{fig:sn1996cr-lc}
demonstrates that the radio emission begins to sharply ``turn on'' in
mid-1996, signaling that \hbox{SN\,1996cr} must have exploded sometime
prior to this. Several narrow-band optical images from the AAT
fortuitously bracket the SN explosion date between 1995-02-28 and
1996-03-16, allowing us to isolate the SN explosion to within a
year. Final confirmation of \hbox{SN\,1996cr} comes in the form of the
optical spectrum, which signals many telltale features of a SN
embedded in dense circumstellar material (CSM).

We note that \hbox{SN\,1996cr}'s position coincides with a small,
powerful H\,{\sc ii} region. However, it is clear from a comparison of
Figs.~\ref{fig:eso_images} and \ref{fig:hst_images} that much of the
H$\alpha$ is due to the unresolved SN-CSM interaction ($\sim$70\%; see also
Table~\ref{tab:data_hst}) and thus is likely to be associated with
photoionized wind material surrounding the progenitor
\citep[e.g.,][]{Fesen1999}. Contamination is likely to 
be negligible for other wavelengths.

The tightest direct temporal constraints arise from comparing the
$R$-band image taken on 1995-02-28 with the narrow-band [S\,II] image
taken on 1996-03-16. As such, we cannot constrain the spectral form
of \hbox{SN\,1996cr} at these early times (i.e., whether it is line- or
continuum-dominated), which make conversions between various bands
somewhat uncertain. For simplicity, we adopt here an intrinsically
flat continuum absorbed by $E(B-V)=2.2$ mags at early times.  This is
likely conservative, considering typical early SNe are intrinsically
quite blue. The extinction was determined using the three narrow-band
discovery-image magnitude constraints and is $\approx$0.8 mags more
than a standard conversion of our best-fitted $N_{\rm H}=8.4
\times 10^{21}$~cm$^{-2}$ from the late-time \hbox{X-ray} spectra;
a factor of $\sim$2 variance is commonly seen between extinction and
$N_{\rm H}$ in our own Galaxy \citep[e.g.,][]{Burstein1978}, so this
discrepancy is probably not necessarily reason for concern, although
it could be indicative of further early-time absorption. We note that
this model should provide adequate conversions to broad-band
magnitudes since early-time SNe spectra are often dominated by strong,
broad lines that blend to form a pseudo-continuum.

Using {\sc synphot} in {\sc iraf} to convert from the narrow
AAT \ion{He}{2}, [\ion{O}{3}], and [\ion{S}{2}] bands, we find an
equivalent discovery magnitude of $V\approx17.8$. At the distance of
the Circinus Galaxy, this equates to an uncorrected absolute magnitude
of $M_{\rm V}\approx-10.2$, or $M_{\rm Vc}\la-15$ when corrected for
the minimum expected extinction. This provides a strong lower limit
since the SN likely went off prior to the discovery images and the
apparent extinction estimated from the discovery magnitudes is
substantially higher.

\begin{figure}
\vspace{0in}
\centerline{
\includegraphics[width=8.5cm]{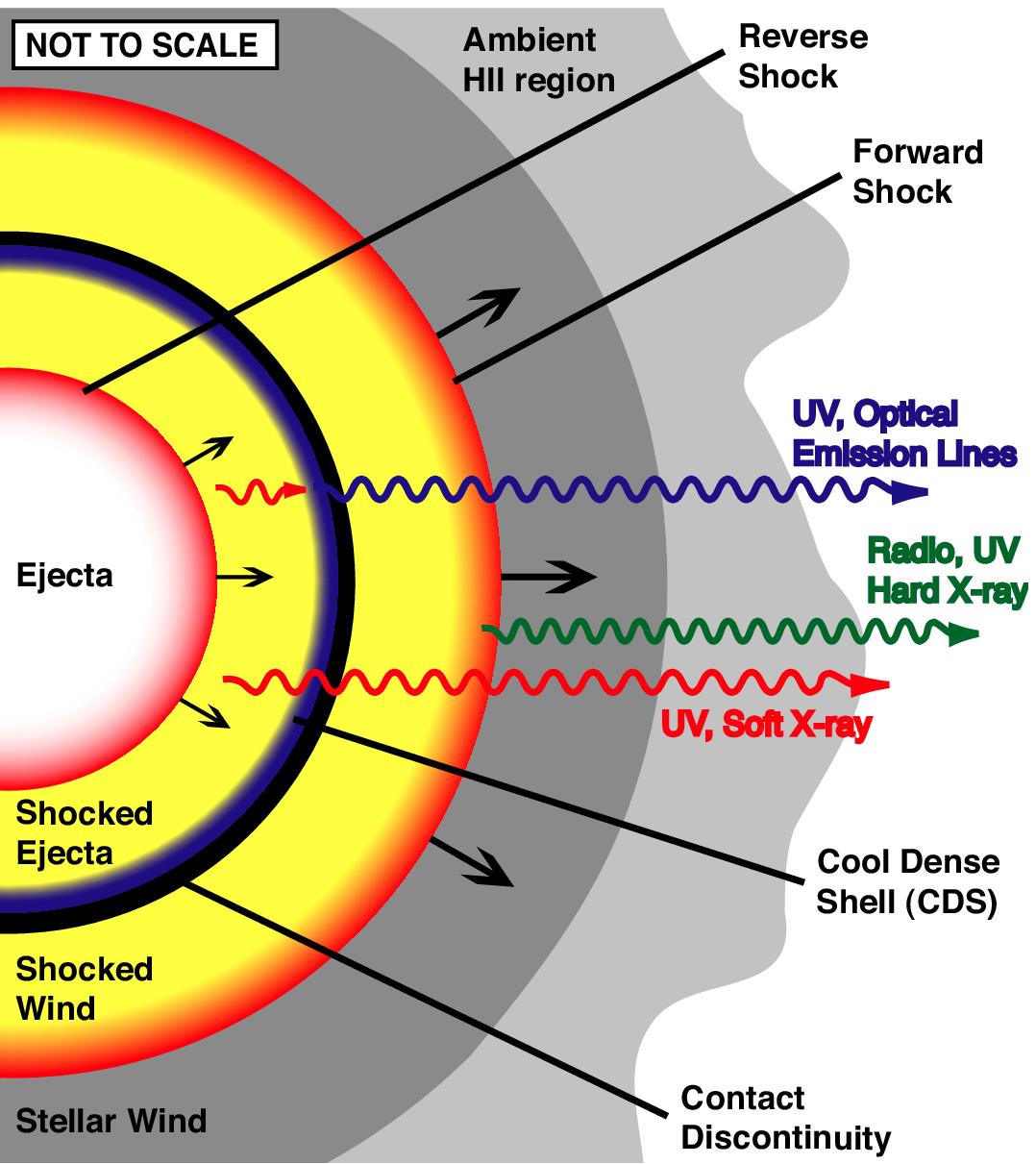}
}
\vspace{0.1in}
\figcaption[sn_pic.eps]{
Idealized one-dimensional cartoon of a SN and its shocks, along with
the stellar wind established CSM, and in the case of \hbox{SN\,1996cr}
a more distant H\,{\sc ii} region the SN is likely to be surrounded by.
The forward shock and shocked CSM are thought to give rise to radio,
UV, and hard X-ray emission ($T\sim10^{9}$~K), while the reverse shock
and shocked ejecta provide the bulk of the soft X-ray emission
($T\sim10^{7}$~K). A CDS immediately behind the contact
discontinuity can potentially absorb the soft X-ray emission, leading
to a variety of photoionized emission lines. Not to scale.
\label{fig:sn_diagram}}
\vspace{0.2in}
\end{figure} 

\section{Characterizing \hbox{SN\,1996cr}}\label{character}

The basic observational picture for core-collapse SNe starts at shock
breakout \citep[e.g.,][hereafter CF94]{Chevalier1994}, where the blast
wave emerges from the progenitor with a typical velocity of
$\ga10^{4}$~km~s$^{-1}$ and propagates into the CSM formed by the
pre-SN stellar wind. The interaction between the ejecta in the forward
shock and the CSM leads to the formation of a reverse shock, which
travels back into the expanding ejecta --- at least relative to the
outward expanding contact discontinuity that separates the shocked
ejecta and the shocked CSM. Fig.~\ref{fig:sn_diagram} provides a
basic one-dimensional picture of the overall scenario. The full
three-dimensional structure could be much more complicated. For instance,
the contact discontinuity is susceptible to Rayleigh-Taylor
instabilities, which are only resolved in
multi-dimensions \citep[e.g.,][]{Chevalier1992,Dwarkadas2000}.
Likewise, asymmetries could exist in the CSM or SN explosion itself
due to the nature of the progenitor or the existence of a binary companion
\citep[e.g., SN\,1979C, SN\,1987A;][]{McCray1993, Crotts2000, Montes2000,
Sugerman2005}. For a typical slow ($v_{w}\sim10$~km~s$^{-1}$), dense
($\mdot \sim 10^{-4}$--$10^{-6}$~M$_{\odot}$~yr$^{-1}$) constant
progenitor wind, the forward shock/CSM interaction produces a very hot
shell of shocked CSM ($T\sim10^{9}$~K), while the reverse shock/ejecta
interaction produces a denser, cooler shell of shocked ejecta
($T\sim10^{7}$~K). The latter interaction has a much higher emission
measure and generates copious \hbox{far-UV} and \hbox{X-ray} emission
which subsequently photoionizes a broad inner ejecta region from which
many of the optical lines are believed to originate. When the CSM
density is high and/or the power-law density distribution of the
ejecta ($\propto R^{-n}$) is steep, the reverse shock will remain
radiative and a relatively thin, cool, dense, partially absorbing
shell (CDS) should form between the reverse shock and the contact
discontinuity. UV and \hbox{X-ray} radiation associated with the
original shock breakout \citep[detected for the first time from
\hbox{SN\,2008D} with {\it Swift};][]{Soderberg2008}, as well as radiation
from the ongoing forward and reverse shocks, may additionally
photoionize the outer CSM. When the CSM is H-rich, this ionization
should give rise to the strong, narrow-line H emission which is
thought to epitomize the
\hbox{type~IIn} class \citep[e.g.,][]{Schlegel1990}. To place 
\hbox{SN\,1996cr} within the context of the above scenario,
we examine our temporal and spectral data in detail.

\subsection{Optical Light Curve}\label{opt_ltcrv}

To assess the early history of \hbox{SN\,1996cr}, we compared the
$V$-band light curves for five \hbox{type~IIn}, as well as 25
additional \hbox{type~II} SNe of varying types, to the optical
constraints detailed in $\S$\ref{data} and plotted in
Fig.~\ref{fig:sn1996cr-lc}.\footnote{SNe light curves were acquired
from \url{http://virtual.sai.msu.su/$^{\sim}$pavlyuk/snlcurve/}.} The
three $V\ga15$ upper limits from R. Evans here provide some additional
insight into the nature of \hbox{SN\,1996cr}'s light curve. We note
that the light curves of \hbox{type~II} SNe are relatively uniform and
can generally be divided into two distinct parts: (1) a sharp rise and
variable decline rate between $\Delta V=0.005$--0.035 per day within
the first $\sim$100~days, followed by (2) a nearly universal decline
from $M_{V}\sim-14$ with a rate of $\Delta V=0.003$--0.01 per day. The
latter is thought to be linked to the decay of
$^{56}$Co \citep{Turatto1990, Patat1994}.

The above parameters define our average template light curve, and we
use as our pivot point the constraints from 1996-03-16 to 1996-03-19,
which lie 0--380 days from maximum. As a consistency check, we note
that the slope between our 1996 and 1999 data points falls roughly
within the expected range for \hbox{type~II} SNe; we adopt $\Delta
V=0.0035$ per day to fit our points. We find that R. Evans' constraint
on 1995-03-31 excludes earlier explosion maximum dates for all but the
faintest observed light curves, while the limit on 1995-08-13 excludes
brighter than average observed light curves up to $\sim$50 days prior
to this date, and the limit on 1996-03-03 fails to exclude any
additional time. The dark grey region in the upper panel of
Fig.~\ref{fig:sn1996cr-lc} shows the range of parameter space which is
still viable, demonstrating that these upper limits
exclude \hbox{SN\,1996cr} from being an extremely
luminous \hbox{type~II} SN over $\sim$15\% of the explosion
window. From this we argue that \hbox{SN\,1996cr} ultimately had an
uncorrected absolute $V$-band magnitude somewhere between $-10.2$ and
$-15.5$. Once corrected for $\ga$4.5 magnitudes of
extinction, \hbox{SN\,1996cr} falls within the typical range for
luminous \hbox{type~II} SNe. At late times, \hbox{SN\,1996cr} lies
well above an extrapolation of the average decay rate, and is in fact
$\sim$1.5 magnitudes brighter in 2006 than in 1999. This fact
demonstrates robustly that \hbox{SN\,1996cr} must be buoyed either by
circumstellar interaction or unresolved light
echoes \citep[e.g.,][]{Patat2005, Patat2006}. We discuss these two
possibilities further in $\S$\ref{opt_spectra_results}.

Several SNe (particularly of the IIn sub-type) have been shown to be
strong near-IR emitters, due to reprocessed emission from hot
dust \citep[e.g.,][]{Gerardy2002, Pozzo2004}. Whether the hot
dust arises from preexisting dust in the circumstellar gas or from
newly formed dust in the ejecta is still unclear. The {\it HST} NIC3
narrow-band constraints we have demonstrate that any near-IR emission
above the earlier ESO near-IR imaging is likely to be minimal. Thus if
a significant hot-dust component formed, it must have either
been short-lived (e.g., destroyed by the initial UV flash from the SN
or rapidly cooled) or occurred after the NIC3 observations in
1998. Mid-IR observations will ultimately constrain the
dust-content of \hbox{SN\,1996cr} better, since the bulk of
anticipated dust is likely to be at temperatures of
$\sim$100~K \citep[the dust temperature in SN\,1987A, for instance, is
$\approx$180~K;][]{Dwek2008}. While the pre-SN optical imaging does
not provide useful constraints on the progenitor due to the strong
extinction toward the Circinus Galaxy, the ESO near-infrared data does
not suffer as much. The $J$-band limits in 1994 June are strong enough
to exclude stars with $M_{\rm J}\la-7.5$, effectively excluding the
bright end of the luminous blue variable (LBV) distribution.

\subsection{Optical Spectrum}\label{opt_spectra_results}

As Fig.~\ref{fig:sn1996cr-spec} demonstrates, the optical spectrum
of \hbox{SN\,1996cr} exhibits strong narrow emission lines of H and O
that typify the \hbox{type~IIn} SNe class \citep[e.g.,][]{Schlegel1990}
superimposed on very broad emission complexes. The strong narrow lines
suggest that circumstellar interaction plays an important role in the
overall emission almost from the onset and, because of the strength of
this emission, it can often mask the true nature of the photospheric
emission.  For instance, \hbox{SN\,1998S} \citep{Fassia2001} was
spectrally a type~IIn and photometrically a type~IIL, \hbox{SN\,2002ic}
\citep{Hamuy2003, Deng2004} and \hbox{SN\,2005gj} \citep{Aldering2006, Prieto2008}
were spectral ``hybrids'' showing type~IIn-like H lines superposed on
otherwise type~Ia-like spectra, and \hbox{SN\,2001em}
\citep{Chugai2006} spectrally evolved from a type~Ic to a
type~IIn on the timescale of a few years.

\begin{figure}
\vspace{-0.1in}
\centerline{
\hglue-0.5cm{\includegraphics[width=9.0cm]{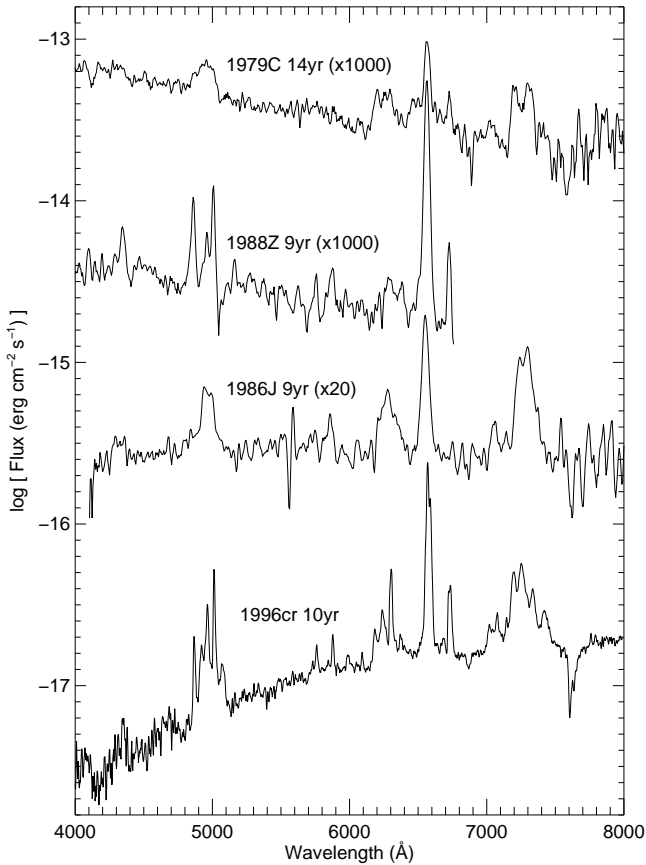}}
}
\vspace{0.0cm} 
\figcaption[sn1996cr_comp.eps]{
A comparison of the optical spectrum of \hbox{SN\,1996cr} with those
for a few other type~IIn supernovae: SN\,1979C \citep{Leibundgut1991},
SN\,1986J \citep{Fesen1999}, SN\,1988Z \citep{Aretxaga1999}. Spectra
have been smoothed over a 3~pixel scale to facilitate easier
comparisons. \hbox{SN\,1996cr} has strong, narrow components similar to
SN\,1988Z as well as broad asymmetric O complexes like SN\,1979C and
SN\,1986J. \label{fig:sn1996cr-compare}}
\vspace{0.5cm} 
\end{figure} 

In the case of \hbox{SN\,1996cr}, where we have limited temporal
coverage during the early phases, we cannot deduce its early
photometric or spectral type. We must rely solely on the late-time SN
emission lines in our VLT spectrum to provide diagnostics on ejecta
abundances and shock emission processes, as well as potential
information on the mass-loss history and evolutionary status of the SN
progenitor \citetext{e.g., CF94, \citealp{Fransson2002},
\citealp{Chevalier2005}}. As Fig.~\ref{fig:sn1996cr-compare} shows,
aside from the heavy extinction, the spectrum of \hbox{SN\,1996cr}
bears many striking similarities to the classic \hbox{type~IIn} SNe,
\hbox{SN\,1979C} \citep{Fesen1999}, \hbox{SN\,1986J}
\citep{Leibundgut1991}, and \hbox{SN\,1988Z} \citep{Aretxaga1999}, and
thus can probably be considered to be representative of the 
class as a whole.

We used the contributed {\sc iraf} package {\sc specfit}
\citep{Kriss1994} to model the various spectral components of
\hbox{SN\,1996cr}. We began by fitting a power-law continuum, absorbed
using the Galactic extinction curve of \citet{Cardelli1989} with
$R_{\rm V}$ fixed at 3.1.\footnote{While there is known to be some
variance in the extinction curves from galaxy to galaxy
\citep[e.g.,][]{Calzetti1994}, such differences should be relatively
minimal above 4000\,\AA. Thus for simplicity we model the extinction
using this particular Galactic extinction model since a significant
fraction of the extinction comes from our own Galaxy.} However, since
there is some degeneracy between the power-law slope and degree of
extinction, we fixed $E(B-V)$ to a value of 1.8 determined by the
narrow-line Balmer decrement\footnote{Note that the decrement has a
slight dependence on temperature and virtually none on density.}  and
then determined a best-fitted power-law slope of 3.4. This single
powerlaw, however, underestimates the curvature of the continuum
between 5500--7000\,\AA~ by $\approx$10\%, and thus conversely leads
to overestimates in some line fluxes and widths. To fit the continuum
more accurately, we instead adopted a broken power-law model with
slopes of 2.65 above and 4.47 below a break at 6400\AA, respectively,
yielding a statistically reliable model of the continuum over the
entire spectrum. Note that the narrow Balmer lines are thought to
arise from the unshocked progenitor wind and thus can be considered as
the outer layer of emission. We therefore consider the extinction
derived from the Balmer decrement to be a lower limit and caution that
some interior components could suffer considerably higher
extinction. We do not, however, see strong evidence of P~Cygni
absorption in any of the emission lines, suggesting that further
absorption in the vicinity of \hbox{SN\,1996cr} is probably
minimal. This may also imply that the bulk of the CSM is fully
ionized. We find only marginal evidence for Na~D absorption
(\hbox{$\sim1.5\sigma$}), which can be used to approximate the overall
extinction \citep[e.g.,][]{Turatto2003}. The strength of this
absorption appears too small compared to our other estimates, although
the spectral resolution, signal-to-noise, and potential for
contamination from complex line emission (see below) in the VLT
spectrum severely compromise any meaningful constraint. We contend
that the lack of broad-band blue and near-UV detections, compared to
the strong blue continua typically seen in comparable SNe \citep[e.g.,
\hbox{SN\,1979C}, \hbox{SN\,1980K}, \hbox{SN\,1988Z};][]{Fesen1999,
Immler2005}, is fully consistent with our estimated extinction
parameters and hence is reasonable. 

We proceeded to fit all of the obvious emission lines with narrow
Gaussian components. The parameters of each line were originally fit
separately, but eventually the central wavelengths and FWHMs of the H,
He, and heavier element lines were tied together as denoted in
Table~\ref{tab:line_ids} to reduce free parameters and improve error
estimates; this was justified by the fact that the free-parameter
values overlapped with the nominal fixed ratio values to within
errors. We note that there were, however, several large broad residual
wings around the H$\alpha$ and O lines. Thus we fit blended narrow and
broad Gaussians, again linking central wavelengths and FWHMs as
denoted in Table~\ref{tab:line_ids}. We describe each element
separately below.

\begin{deluxetable*}{lllllll}
\tabletypesize{\scriptsize}
\tablewidth{0pt}
\tablecaption{Line Identifications\label{tab:line_ids}}
\tablehead{
\colhead{\#} &
\colhead{Line} &
\colhead{Component} &
\colhead{$F$} &
\colhead{$F^{c}$} &
\colhead{$\lambda_{c}$} &
\colhead{FWHM}
}
\tableheadfrac{0.05}
\startdata
1  & [\ion{O}{2}] $\lambda$7319      & n  &   1.4$\pm$ 0.5 &   49.6$\pm$17.2 & 7324.9$\pm$1.4  & $=$29 \\
2  &                                 & b  &  13.6$\pm$ 1.0 &  496.0$\pm$34.8 & 7324.9~($=$1*)  & $=$29 \\
3  &                                 & c1 &   6.2$\pm$ 0.5 &  208.0$\pm$16.4 & 7420.1~($=$30*) & $=$30 \\
4  &                                 & c2 &  19.3$\pm$ 0.7 &  746.0$\pm$26.8 & 7251.1~($=$31*) & $=$30 \\
5  &                                 & c3 &  16.8$\pm$ 0.5 &  681.8$\pm$20.7 & 7190.5~($=$32*) & $=$30 \\
6  & [\ion{Ar}{3}] $\lambda$7136     & n  &   1.8$\pm$ 0.2 &   76.4$\pm$ 9.7 & 7136.8~(f)      & $=$8 \\
7  & \ion{He}{1} $\lambda$7065       & n  &   3.1$\pm$ 0.3 &  138.3$\pm$12.7 & 7065.7~($=$21*) & $=$21 \\
8  & [\ion{S}{2}] $\lambda$6731      & n  &   3.6$\pm$ 0.3 &  210.1$\pm$18.9 & 6732.6$\pm$0.8  & 741$\pm$24 \\
9  & [\ion{S}{2}] $\lambda$6716      & n  &   4.0$\pm$ 0.1 &  239.0$\pm$ 7.8 & 6717.6~($=$8*)  & $=$8 \\
10 & \ion{He}{1} $\lambda$6678       & n  &   1.1$\pm$ 0.2 &   70.0$\pm$13.8 & 6675.7~($=$21*) & $=$21 \\
11 & [\ion{N}{2}] $\lambda$6583      & n  &  19.9$\pm$ 0.7 & 1310.0$\pm$45.4 & 6582.8$\pm$0.2  & $=$8 \\
12 & H$\alpha$                       & n  &  34.5$\pm$ 0.8 & 2306.8$\pm$51.8 & 6563.0$\pm$0.2  & 669$\pm$18 \\
13 &                                 & b  &  9.5$\pm$ 1.0 &  634.4$\pm$65.2 & 6563.0~($=$12)  & 4077$\pm$398 \\
14 & [\ion{N}{2}] $\lambda$6548      & n  &   6.4$\pm$ 0.0 &  432.3$\pm$ 0.0 & 6547.8~($=$11*) & $=$8 \\
15 & [\ion{O}{1}] $\lambda$6300      & n  &   5.9$\pm$ 0.4 &  486.7$\pm$30.7 & 6296.2$\pm$0.4  & $=$8 \\
16 &                                 & b  &   5.3$\pm$ 0.6 &  434.9$\pm$51.3 & 6296.2~($=$15)  & $=$29 \\
17 &                                 & c1 &   2.1$\pm$ 0.3 &  159.8$\pm$26.0 & 6376.8~($=$30*) & $=$30 \\
18 &                                 & c2 &   5.8$\pm$ 0.4 &  504.3$\pm$35.5 & 6231.6~($=$31*) & $=$30 \\
19 &                                 & c3 &   2.9$\pm$ 0.3 &  263.7$\pm$29.2 & 6179.6~($=$32*) & $=$30 \\
20 & [\ion{Fe}{7}] $\lambda$6087     & n  &   0.5$\pm$ 0.2 &   43.8$\pm$16.3 & 6087.0~(f)      & $=$8 \\
21 & \ion{He}{1} $\lambda$5876       & n  &   1.6$\pm$ 0.2 &  187.8$\pm$24.1 & 5873.9$\pm$0.8  & 913$\pm$65 \\
22 & [\ion{N}{2}] $\lambda$5755      & n  &   1.2$\pm$ 0.2 &  152.3$\pm$22.2 & 5755.0~(f)      & $=$8 \\
23 & [\ion{O}{1}] $\lambda$5577      & n  &   0.2$\pm$ 0.1 &  35.6$\pm$15.4 & 5579.0~($=$28*) & $=$8 \\
24 &                                 & b  &   0.3$\pm$ 0.2 &  49.3$\pm$28.5 & 5579.0~($=$28*) & $=$29 \\
25 &                                 & c1 &   0.3$\pm$ 0.2 &  48.5$\pm$26.1 & 5644.5~($=$30*) & $=$30 \\
26 &                                 & c2 &   0.3$\pm$ 0.1 &  49.3$\pm$20.8 & 5516.0~($=$31*) & $=$30 \\
27 &                                 & c3 &   0.3$\pm$ 0.2 &  48.2$\pm$31.0 & 5469.9~($=$32*) & $=$30 \\
28 & [\ion{O}{3}] $\lambda$5007      & n  &   5.2$\pm$ 0.3 &1638.6$\pm$92.3 & 5007.9$\pm$0.3  & $=$8 \\
29 &                                 & b  &   3.5$\pm$ 0.4 &1099.1$\pm$130.3 & 5007.9~($=$28)  & 2795$\pm$174 \\
30 &                                 & c1 &   2.4$\pm$ 0.3 & 682.5$\pm$71.4 & 5066.4$\pm$1.1  & 2049$\pm$19 \\
31 &                                 & c2 &   4.5$\pm$ 0.3 &1543.0$\pm$117.1 & 4952.6$\pm$0.6  & $=$30 \\
32 &                                 & c3 &   1.9$\pm$ 0.3 & 690.5$\pm$94.2 & 4910.8$\pm$0.6  & $=$30 \\
33 & [\ion{O}{3}] $\lambda$4959      & n  &   1.6~($=$27*) & 540.8~($=$27*) & 4960.0~($=$28*) & $=$28 \\
34 &                                 & b  &   1.1~($=$28*) & 362.7~($=$28*) & 4960.0~($=$28*) & $=$29 \\
35 &                                 & c1 &   0.7~($=$29*) & 225.2~($=$29*) & 5018.9!($=$30*) & $=$30 \\
36 &                                 & c2 &   1.1~($=$30*) & 509.2~($=$30*) & 4905.2~($=$31*) & $=$30 \\
37 &                                 & c3 &   0.6~($=$31*) & 227.9~($=$31*) & 4866.8~($=$32*) & $=$30 \\
38 & H$\beta$                        & n  &   1.9~($=$11*) & 745.0~($=$11*) & 4861.1~($=$12*) & $=$12 \\
39 & \ion{He}{2} $\lambda$4685       & n  &   0.4$\pm$0.1  & 221.8$\pm$59.6 & 4684.5~($=$21*) & $=$20 \\
40 & H$\gamma$                       & n  &   0.4~($=$11*) & 350.2~($=$11*) & 4340.0~($=$12*) & $=$12 \\
\enddata
\tablecomments{
{\it Column 1:} Component number.
{\it Column 2:} Emission line.
{\it Column 3:} Component: Narrow (n), Broad (b), Complex (c\#).
{\it Column 4:} Observed flux in units of
$10^{-16}$~ergs~s$^{-1}$~cm$^{-2}$. ``*'' indicates the flux of the
line is tied to another line according to the relative ratio of their
respective atomic line intensities.
{\it Column 5:} Extinction-corrected flux in units of
$10^{-16}$~ergs~s$^{-1}$~cm$^{-2}$. ``*'' indicates the flux of the
line is tied to another line according to the relative ratio of their
respective atomic line intensities.
{\it Column 6:} Central wavelength in \AA. ``*'' indicates the central
wavelength of the line is tied to another line according to the ratio
of their respective atomic rest wavelengths.
{\it Column 7:} Gaussian Full-Width Half-Maximum in km~s$^{-1}$.
}
\end{deluxetable*}

\begin{deluxetable*}{lllllll}
\tabletypesize{\scriptsize}
\tablewidth{0pt}
\tablecaption{Line Ratios\label{tab:ratios}}
\tablehead{
\multicolumn{2}{c}{Line Ratio} &
\colhead{``n''} &
\colhead{``b''} &
\colhead{``c1''} &
\colhead{``c2''} &
\colhead{``c3''}
}
\tableheadfrac{0.05}
\startdata
$[$\ion{O}{3}$]$ & $I(4959+5007)/I(4363)$      & $>5.40$ & $>3.62$ & $>5.08$ & $>2.25$ & $>2.28$ \\ 
$[$\ion{O}{2}$]$ & $I(3726+3729)/I(7319+7331)$ & $<703$ & $<3.02$ & $<1.78$ & $<7.46$ & $<1.90$ \\ 
\ion{O}{1}/$[$\ion{O}{2}$]$ & $I(7774)/I(7319+7331)$      & $<0.50$ & $<0.16$ & $<0.12$ & $<0.48$ & $<0.34$ \\ 
$[$\ion{O}{1}$]$ & $I(6300+6364)/I(5577)$      & $>3.95$ & $>1.90$ & $>3.14$ & $>1.07$ & $>1.63$ \\ 
$[$\ion{S}{2}$]$ & $I(6731)/I(6716)$           & $0.88^{+0.18}_{-0.15}$ & & & & \\ 
$[$\ion{S}{2}$]$ & $I(6716+6731)/I(4068+4076)$ & $>0.31$ & & & & \\ 
$[$\ion{N}{2}$]$ & $I(6548+6583)/I(5755)$      & $11.44^{+9.54}_{-3.75}$ & & & & \\ 
$[$\ion{Ar}{3}$]$ & $I(7136+7751)/I(5192)$     & $>0.84$ & & & & \\ 
 \ion{He}{1} & $I(7065)/I(5876)$            & $0.74^{+0.78}_{-0.35}$ & & & & \\ 
 \ion{He}{1} & $I(7065)/I(6678)$            & $1.98^{+4.15}_{-1.07}$ & & & & \\ 
 \ion{He}{1} & $I(6678)/I(5876)$            & $0.37^{+0.59}_{-0.26}$ & & & & \\ 
\enddata
\tablecomments{
{\it Column 1:} Line-intensity ratio.
{\it Columns 2-6:} Fitted component values.
}
\end{deluxetable*}

We should caution that the late-time spectrum we observe
for \hbox{SN\,1996cr} could potentially be contaminated by early-time
SN spectral features due to the presence of unresolved light echoes
from dust in the immediate circumstellar environment or in the
intervening interstellar
medium \citep[e.g.,][]{Patat2005,Patat2006}. Notably, the spectra of
SNe near maximum light are the most likely to continue on as light
echoes, and thus we would expect such contamination to be isolated
primarily to the continuum of \hbox{SN\,1996cr} since early-time
spectra of SNe routinely exhibit either blue continua or
psuedo-continua comprised of extremely broad emission lines which
trace the high velocity ejecta \citep{Filippenko1997}.
For \hbox{SN\,1996cr}, we find only distinct emission lines of
low-to-moderate velocity (i.e., $<5,000$~km~s$^{-1}$) on top of a
smooth continuum, with the only notable exception being the region
immediately surrounding He~$\lambda$5876. The equivalent widths of
these emission lines are also comparable to other well-studied SNe
such as those in Fig.~\ref{fig:sn1996cr-compare}, suggesting at least
that SN\,1996cr is not unusual. We additionally included a few
early-time light curves of typical SNe in our spectral fitting, and
found that aside from some degeneracy with our continuum model, such
spectra are not generally compatible. Thus we argue that the majority
of our line estimates, with the possible exception of H$\alpha$, are
unlikely to be strongly contaminated. In the case of H$\alpha$, there
are rare exceptions, such as the \hbox{type~IIn} SN\,1994W and
SN\,1994Y, where early spectra display bright, relatively narrow H
lines on top of blue continua from the
outset \citep[e.g.,][]{Wang1996, Chugai2004}. In these cases, the
narrow lines are interpreted as arising from the SN shock transmitted
into a dense, clumpy CSM very close in to the SN
($\sim$10$^{15}$~cm). Our radio and X-ray constraints at early times
suggest this is an unlikely scenario for SN\,1996cr (see
$\S$\ref{radio_ltcrv} and $\S$\ref{xray_ltcrv}). However, since we
lack early optical spectra of \hbox{SN\,1996cr}, interpretation of the
H$\alpha$ emission should be viewed with some discretion.

\begin{figure}
\vspace{-0.05in}
\centerline{
\hglue-0.2cm{\includegraphics[width=9.0cm]{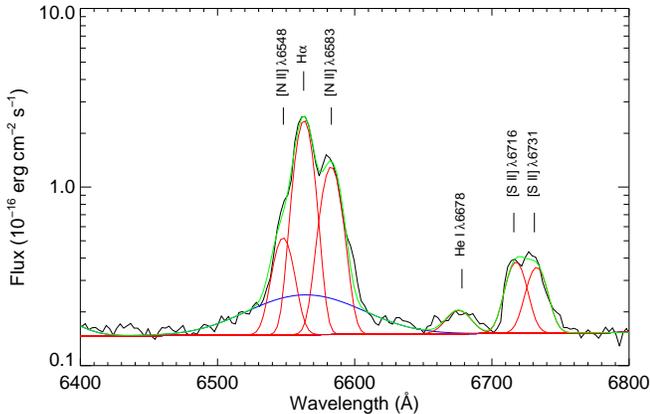}}
}
\vspace{-0.1cm} 
\figcaption[sn1996cr_hacomplex.eps]{
%
%
A section of the spectrum in Fig.~\ref{fig:sn1996cr-spec} showing in
detail the fitted components in the vicinity of the H$\alpha$
line. Obvious emission lines have been identified and an empirical
model has been fit to the data as described in
$\S$~\ref{opt_spectra_results}; again the green line shows the overall
fit to the spectrum while the red (blue) lines show the individual
narrow (broad) components. Although the narrow H$\alpha$ strongly
overlaps with the \ion{N}{2} doublet, our spectral resolution is
sufficient to deblend it. \label{fig:sn1996cr-hacomplex}}
%
%
\vspace{0.1cm} 
\end{figure} 

\subsubsection{Hydrogen}\label{hydrogen_spectra_results}

We find \hbox{SN\,1996cr} exhibits both narrow (669$\pm$18~km~s$^{-1}$)
and broad (4077$\pm$398~km~s$^{-1}$) components of H$\alpha$, with
fluxes of (2.31$\pm$0.05)$\times10^{-13}$~ergs~s$^{-1}$~cm$^{-2}$ and
(6.34$\pm$0.65)$\times10^{-14}$~ergs~s$^{-1}$~cm$^{-2}$, respectively.
This portion of the spectrum is shown in detail in
Fig.~\ref{fig:sn1996cr-hacomplex}. The narrow line appears to be
marginally resolved above our estimated instrumental resolution,
although this fact should be regarded with some caution since the line
is strongly blended with \ion{N}{2}; we await confirmation via
high-resolution spectroscopic follow-up. 

The flux found in our VLT spectrum is consistent to within
errors with that measured from the {\it HST} F656N filter seven years
earlier, indicating that the overlapping H\,{\sc ii} region still only
contributes at most 30\% to the total narrow-line flux above and 10\%
to the broad-line flux. Accounting for this contamination, we find
corresponding extinction-corrected luminosities of
(2.79$\pm$0.06)$\times10^{38}$~ergs~s$^{-1}$ and
(9.86$\pm$0.10)$\times10^{37}$~ergs~s$^{-1}$ for the narrow and broad
components, respectively. The energy generated in either the narrow or
broad H$\alpha$ lines is significantly above that predicted by the
decay of $^{56}$Co--$^{56}$Fe at this late stage
\citep[e.g.,][]{Chugai1991}, and therefore must instead be related either to
mechanical energy associated with the SN ejecta-wind interaction or
potential light echoes.

The line width and systemic velocity of the narrow component argue for
an origin beyond the blast wave, associated with either the shocked or
unperturbed progenitor wind \citep[e.g.,][]{Fransson2002}. If the line
does arise from a dense, pre-SN, H-rich CSM, it must have been
photoionized and heated by UV and \hbox{X-ray} radiation from the
massive progenitor and the initial shock breakout of the SN. The fact
that the H$\alpha$ luminosity has remained relatively constant between
the epochs of the VLT and {\it HST} observations suggests that this
region is now being sustained by ionizing flux from the current
ejecta-wind shock zone (see $\S$\ref{xray_ltcrv}), although again
contaminating echoes from prior epochs is a concern here. For
comparison, the H$\alpha$ flux of other \hbox{type~IIn} SNe such
as \hbox{SN\,1988Z}, \hbox{SN\,1995G}, and \hbox{SN\,1995N} decreased by
at least an order of magnitude in $\sim$4~yrs \citep{Aretxaga1999,
Pastorello2002, Fransson2002}. The H$\alpha$ luminosity thus provides
an important diagnostic on the total ionizing radiation that
complements the X-ray observations, as this line arises as a result of
recombination and collisional excitation. If this H$\alpha$ component
is associated with CSM interaction and is powered by photoionization,
then the X-ray luminosity from the forward and reverse shocks should
roughly be proportional to the H$\alpha$ luminosity following Eq. 3.7
in CF94 such that

\begin{equation}
L({\rm H}\alpha)\approx4.4\times10^{37} (T/10^{7} {\rm K})^{-1}
(L_{\rm X}/10^{40})~{\rm ergs}~{\rm s}^{-1}.
\end{equation}

\noindent This implies $L_{X}=7.0\times10^{40}(T/10^{7} K)$~ergs~s$^{-1}$, 
however, which is two orders of magnitude larger than the product of
observed absorption-corrected \hbox{X-ray} luminosity and
temperature. This would argue for the soft X-rays from the
reverse shock, which should dominate bolometrically, to perhaps be
radiative and hence reprocessed by either unshocked low-density ejecta
or high-density material in the CDS (this possibility is discussed
further in $\S$\ref{oxygen_spectra_results}).  Alternatively, the
narrow line could arise from shocks being driven into dense,
slow-moving clumps associated with the progenitor wind. This would
more naturally explain the discrepancy in observed and estimated
ionizing flux, in which case the line is instead powered by
collisional excitation.

If the narrow H$\alpha$ stems from the stellar wind, then the
H$\alpha$ luminosity can also be related to the mass of the unshocked
wind as
 
\begin{equation}
L({\rm H}\alpha)=(1-5)\times10^{39} (\mdot_{-4}/v_{w,1})^{2}~r_{18}^{-1}\ {\rm ergs}\ {\rm s}^{-1},
\end{equation}

\noindent where $r_{18}$ is the radius in units of
$10^{18}$~cm, $v_{w}$ is the wind velocity in units of 10~km~s$^{-1}$,
and $\mdot_{-4}$ is the mass-loss rate in units of
$10^{-4}$~M$_{\odot}$~yr$^{-1}$ \citep{Deng2004, Wang2004}. We assume
an emissivity based on Case B recombination at a temperature of
10,000~K, which is in rough agreement with our extinction-corrected
H$\alpha$/H$\beta$/H$\gamma$ ratios. Our H$\alpha$ luminosity implies
a mass-loss rate of \hbox{$\sim$$10^{-5}\, v_{w,1}\,
r_{18}^{0.5}$~M$_{\odot}$~yr$^{-1}$}.  For a typical red supergiant
lifetime of $\ga10^{5}$~yr, this implies that at least a few M$_{\odot}$
of hydrogen-rich matter was cast off into the CSM.

The high-velocity component of H$\alpha$, on the other hand, must be
related directly to the blast wave in order to have achieved its
present dispersion. This component could be associated with the
reverse shock or CDS behind the reverse shock, perhaps from swept-up
wind material or hydrogen that remained on the surface of the
progenitor. The broad-line H$\alpha$ velocity is thought to provide a
lower limit to the true shock velocity. However, an observed disperion
of $\sim$4,000~km~s$^{-1}$ is much lower than the
15,000--40,000~km~s$^{-1}$ initially expected for a typical
core-collapse SN, and implies that the shock in \hbox{SN\,1996cr} has
already slowed down substantially due to circumstellar
interaction. This observed velocity decrease of at least $\ga$4--10
implies a circumstellar density increase of $\ga$16--100. This line
is almost certainly powered by collisional excitation associated with
shocked ejecta.

We additionally see significant narrow-line emission in H$\beta$ and
H$\gamma$, as well as marginally in H$\delta$. Broad emission
components could also exist for these lines; however, for typical
line ratios with respect to H$\alpha$ they would be marginal at
best. Additionally, such emission would be impossible to deblend from
the [\ion{O}{3}] emission complexes.

\begin{figure}
\vspace{-0.1in}
\centerline{
\includegraphics[width=9.0cm]{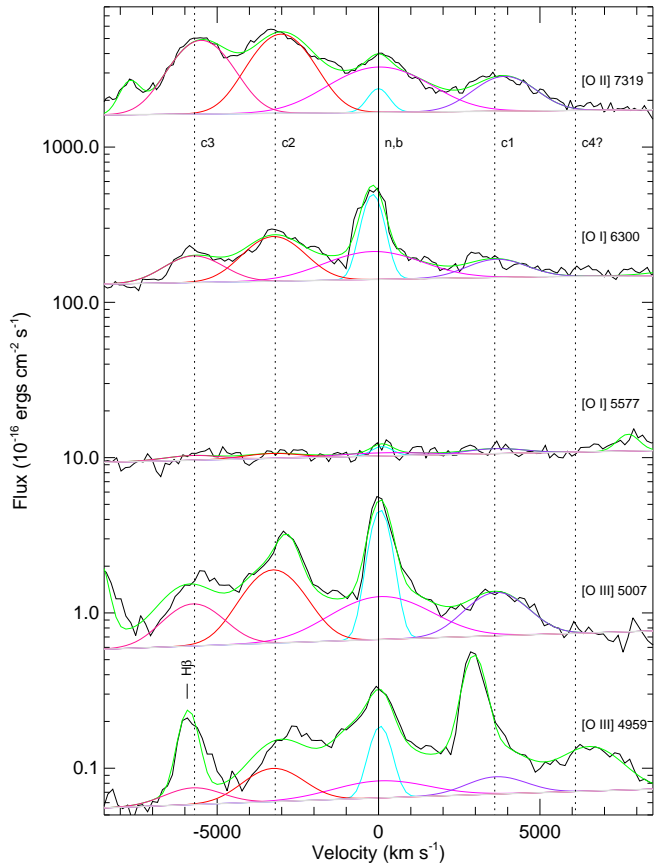}
}
\figcaption[sn1996cr_complex.eps]{
%
%
Velocity structure of O emission lines in units of km~s$^{-1}$. The
VLT spectra in the vicinity of five labeled complexes (black) and
their respective model fits (green showing the overall full component
fit, while contributions from a distinct velocity component are shown
in different colors) are shown. Each spectrum is offset by a factor of
10 for clarity. The solid vertical line denotes the approximate
position of the stationary narrow (cyan) and broad (magenta)
components, while the dashed vertical lines denote the expansion
velocities of components c1 (purple), c2 (red), and c3 (pink); a
potential symmetric component c4 is not evident in the data. Although
several of the lines are strongly blended, each of the components
remains relatively distinct. The line components for $\lambda$5007 and
$\lambda$4959 overlap such that in a few instances no single component
obviously dominates.
%
%
\label{fig:sn1996cr-complex}}
\vspace{0.2cm} 
\end{figure} 

\subsubsection{Oxygen}\label{oxygen_spectra_results}

We find evidence for one narrow-line and four broad-line velocity
components in the profiles of the
[\ion{O}{3}]$\lambda\lambda$5007,4959, [\ion{O}{1}]$\lambda$6300, and
[\ion{O}{2}]$\lambda$7319 lines as shown in
Fig.~\ref{fig:sn1996cr-complex}, with velocities of $3577\pm65$
(``c1''), $66\pm18$ (``n'', ``b''), $-3247\pm36$ (``c2''), and
$-5755\pm37$~km~s$^{-1}$ (``c3''). We note there is also marginal
evidence for similar multi-component emission complexes associated
with the [\ion{O}{1}]$\lambda$5577 and [\ion{O}{3}]$\lambda$4363
lines. This is somewhat in contrast to the late-time spectra of
many type~II SNe such as \hbox{SN\,1993J}, which show evidence for
more irregular, ``clumpy'', newly synthesized O
complexes \citep{Matheson2000}.

A few type~II SNe have shown evidence for three-peaked profiles, which
have either been ascribed to inhomogeneous, partially absorbed disk-
or torus-shaped CSMs \citep[e.g.,][]{Gerardy2000, Leonard2000}, or
from line formation in the CDS \citep[e.g.,][]{Fransson2005}. In such
cases, however, the profiles arise not from the O lines, but from
H$\alpha$, \ion{Mg}{2}, or Ly$\alpha$; this may indicate
that \hbox{SN\,1996cr} is O-rich. Nonetheless, the first scenario
typically yields two relatively symmetric positive and negative
velocity peaks indicative of emission from an equatorial ring, while a
third central peak could emerge due to a population of dense clouds
embedded in a much less dense CSM. In the second scenario, the CDS can
only lead to such peaks under a very specific set of conditions such
that it is very optically thick to such line emission, it is thin
enough that the velocity gradient over the shell is small compared
to the thermal velocity, and finally it has a macroscopic velocity
larger than its thermal velocity. The CDS could be clumpy, lowering
the filling factor and allowing some emission to escape. A final
possibility for this velocity structure could arise from the SN blast
wave impacting a dense shell produced by a wind-blown bubble. In this
case, several secondary shocks could separate the CSM into multiple
concentric shells \citep[e.g.,][]{Dwarkadas2005}, such that each peak
may represent a different velocity shell. The inner (red-shifted)
shells would be preferentially absorbed compared to the outer
(blue-shifted) shells, just as we observe.

For \hbox{SN\,1996cr}, we see a broad central peak (``b'',
FWHM$=$2795$\pm$174~km~s$^{-1}$) and two roughly symmetric positive
and negative peaks (``c1'' and ``c2'',
FWHM$=2049\pm19$~km~s$^{-1}$). The red peak is always fainter than the
blue one, implying significant internal extinction. Interestingly, we
also see a fourth, faster negative component (``c3'') which does not
have an obvious positive counterpart; with our current sensitivity
such a symmetric component (i.e., ``c4?'')  must be at least a factor
of 2.5 fainter (\hbox{3$\sigma$}), implying significant additional
internal absorption above that of the lower velocity peaks, if it
exists. This ``c3'' component could be indicative of further asymmetry
in the structure of \hbox{SN\,1996cr}. We need only look to the
nearby \hbox{SN\,1987A}, with its triple ring structure, for an
example of such potential complexity \citep{McCray1993, McCray2007}.
While the properties of the CSM of \hbox{SN\,1996cr} appear to differ
dramatically from those of \hbox{SN\,1987A}, the spectral features
hint at an environment that is equally dynamic and exotic. 
Another possibility is that the ``c3'' component represents a light
echo from an earlier epoch, where the O lines were shifted by a larger
velocity and more heavily absorbed. We reject this, however since it
would require the ``c3'' oxygen features to peak strongly in the blue
part of the spectrum, while we detect them clearly throughout the
spectrum.

Notably, when we apply our five-component model to the O lines, we
find that our constraints on [\ion{O}{1}]$\lambda$6363 are consistent
with zero flux if we fit [\ion{O}{1}]$\lambda$6363 and
[\ion{O}{1}]$\lambda$6300 separately. Given the degeneracy between the
potential broad components of these two lines though, this only
provides weak $I(\lambda6363)/I(\lambda6300)$ constraints for the
various components (n:$<1.0$, b:$<0.8$, c1$<0.6$, c2$<1.5$,
c3$<0.9$). Such ratios are all fully consistent with the normal
transition probability of 0.3. For the following, we fit only
$\lambda$6300 and use its flux to estimate the total intensity
$I(\lambda6300+\lambda6363)$ for the various components. As such we
may incur some small additional errors associated with flux
contamination between different velocity components of the two lines.
We also caution that the complex which we attribute solely to 
[\ion{O}{2}]$\lambda$7331 is likely in fact to be a blend of
[\ion{O}{2}]$\lambda\lambda$7319,7331 and
[\ion{Ca}{2}]$\lambda\lambda$7291,7325. Using our five-component
model, however, we find that additional lines do not significantly
improve $\chi^{2}$. The relative symmetry and constraints on line
centers of the individual components imply that other lines do not
contribute strongly to the overall emission, although notable small
systematic residuals suggest minor blending is likely. The overlapping
broad-line O profiles seen in Fig.~\ref{fig:sn1996cr-complex} appear
to be overestimating the valleys in between distinct component peaks,
indicating that either the profiles are not represented well by
Gaussians or that P~Cygni-like absorption may be present but
masked by the overall complexity of the emission. Higher-resolution
spectroscopy is needed to place better constraints on these potential
emission and absorption contributions.

The fact that the various components show up in all of the O lines
strongly argues for a scenario in which they arise from physically
associated regions within the SN. Since we see several ionization
states though, it is likely that there is strong stratification within
these distinct spatial regions (very likely due to an extended
partially-ionized zone; e.g., CF94) and it is only appropriate to
employ nebular-to-auroral line ratios for a given ionic species to
estimate physical conditions. However, our constraints on
$\lambda\lambda3726,3729$, $\lambda4363$, and $\lambda5577$ are poor
and do not provide physically interesting temperature and density
limits. Table~\ref{tab:ratios} lists these oxygen ratio constraints.
Notably, the relative observed strengths of the O lines are broadly
consistent with the model predictions of CF94 (see Figs. 4 and 5),
whereby the modeled ejecta is photoionized by X-rays from the
circumstellar interaction.

Finally, there appears to be no sign of the \ion{O}{1}~$\lambda$7774
recombination line either at its systemic value or other velocities
seen in the [\ion{O}{1}] lines (although telluric absorption does
hinder assessment of the blue components). This suggests that the
narrow O resides in a physically distinct region separate from the
narrow H$\alpha$, as one typically sees a $10^{-3}$ ratio.  The broad,
high-velocity O emission is not mirrored in any other elements,
although there is possibly some complexity to the He emission (see
below), further arguing that it probably arises from an O-rich shell
of processed gas associated with freely expanding ejecta.

\subsubsection{Helium}\label{helium_spectra_results}

SN\,1996cr shows evidence for four prominent, narrow He lines
(906$\pm$101~km~s$^{-1}$): He~I$\lambda$7065,
\ion{He}{1}~$\lambda$6678, \ion{He}{1}~$\lambda$5876, and
\ion{He}{2}~$\lambda$4686. The narrow-line width is marginally broader
than that of H$\alpha$, hinting that it may be spatially
distinct. The \ion{He}{1} line ratios do not strongly constrain the electron
temperature and density of the gas. The emissivities of
\citet{Porter2005} constrain the He gas to be at $T\ga10^{4}$~K and
$n_{e}\ga10^{4}$~cm$^{-3}$.

We see substantial broad residual emission around $\lambda$7065 and
$\lambda$5876 which currently defies identification. Unlike the
well-formed velocity components associated with the various O lines,
there appears to be no spectral consistency between different He
lines. For instance, the emission around $\lambda$7065 appears
box-like and quite strong relative to the systemic narrow line, while
the emission around $\lambda$5876 appears more spread out and much
fainter relative to the systemic narrow line. Thus if this broad
emission is associated with He, it is likely present in several
regions of the SN, and may be far more complex than H or O are. There
may be additional marginal broad features around $\lambda$4686,
while any potential features around $\lambda$6678 are masked by the
strong, broad H$\alpha$ emission.

The narrow He emission likely arises from the same regions of the
progenitor wind as the narrow hydrogen emission, whereas the broad
unidentified emission, if it comes from helium, could be swept up
material or expelled clumps. Either way, its complex profile suggests
it is much more irregularly distributed compared to H.

\subsubsection{Nitrogen}\label{nitrogen_spectra_results}

We find evidence for strong, narrow N emission in the spectrum of
\hbox{SN\,1996cr} in the form of both the nebular 
[\ion{N}{2}]~$\lambda\lambda$6583,6548 (which appear as a resolvable
blend with H$\alpha$) and the auroral [\ion{N}{2}]$\lambda$5755
lines. The lines have a best-fitted FWHM of 741$\pm$24~km~s$^{-1}$.
The line-intensity ratio $I(\lambda6583+\lambda6548)/I(\lambda5755)$
implies an electron density of $n_{e}\ga1\times10^5 \times
(10^{4}/T)^{0.5}$~cm$^{-3}$ for $T\la 20,000$~K \citep{Osterbrock1989,
Keenan2001}. The high density suggests that clumping of N is
likely. As argued for the narrow H emission, the majority of the
emission must arise from the SN-CSM interaction given the pre-SN imaging
constraints. The implied velocity of the lines suggest it is
likely to have a circumstellar origin, while the strength of the
emission hints at potential nitrogen enrichment, as has been notably
seen in several other \hbox{type~II} SNe such
as \hbox{SN\,1979C}, \hbox{SN\,1987A}, \hbox{SN\,1993J}, \hbox{SN\,1995N},
and \hbox{SN\,1998S} \citep{Fransson1984, Fransson2002,
Fransson2005}.

\subsubsection{Sulfur}\label{sulfur_spectra_results}

The spectrum shows strong, narrow
[\ion{S}{2}]$\lambda\lambda$6731,6716 emission lines with a
best-fitted FWHM of 741$\pm$24~km~s$^{-1}$, which again clearly must
come from the SN-CSM iteraction given the pre-SN imaging
constraints. This feature is relatively uncommon in late-time type~II
SNe spectra, although it is seen in several comparatively-aged
type~IIn SNe spectra such as \hbox{SN\,1988Z} and
\hbox{SN\,1979C} (see Fig.~\ref{fig:sn1996cr-compare}). The
[\ion{S}{2}] line-intensity ratio $I(\lambda 6731)/I(\lambda 6716)$
implies an electron density of $n_{e}=400^{+100}_{-60} \times
(10^{4}/T)^{0.5}$~cm$^{-3}$
\citep{Osterbrock1989}. We additionally see a hint of the
[\ion{S}{2}]$\lambda\lambda$4076,4069 doublet, although the
significance of this feature is marginal and strongly affected by our
estimated extinction value. Given the strong density contrasts, this
component should reside in a different region from [\ion{N}{2}].

\subsubsection{Argon}\label{argon_spectra_results}

Narrow Ar emission is present in the spectrum of
\hbox{SN\,1996cr} in the form of both the nebular 
[\ion{Ar}{3}]$\lambda\lambda$7751,7135 and the auroral
[\ion{Ar}{3}]$\lambda$5192 lines. Our lower limit to the line
intensity ratio $I(\lambda5192)/I(\lambda 7751+\lambda 7135)$,
however, is too loose to place any useful constraint on the electron
temperature or density of this component \citep[e.g.,][]{Keenan1988}.

\subsubsection{Iron}\label{iron_spectra_results}

We find significant high-ionization
[\ion{Fe}{7}]$\lambda\lambda$6087,5721 lines, but only a marginal
[\ion{Fe}{7}]$\lambda$5158 line. Additionally,
[\ion{Fe}{10}]$\lambda$6375 may be present, although it is currently
lost amid the broad [\ion{O}{1}] emission. The lack of a strong
$\lambda$5158 line constraint unfortunately leads to physically
uninteresting electron-density limits \citep[e.g.,][]{Keenan1987,
Keenan1991}.

\subsubsection{Unidentified}\label{noid_spectra_results}

We also see tentative features at $\lambda$5840, $\lambda$5980,
$\lambda$7011, and $\lambda$7785 which remain unidentified.


\subsection{Radio Light Curve}\label{radio_ltcrv}

An expectation from the interaction between the blast wave and the CSM
is the production of copious non-thermal synchrotron emission from
relativistic electrons and enhanced magnetic fields within the thin
shell trailing the forward shock. The radio emission should be
proportional to the injection spectrum of electrons and various loss
mechanisms, which are effectively determined by the composition and
density of the CSM. At early times, however, the radio emission can be
absorbed via free-free (FF) or synchrotron self-absorption (SSA)
depending on the magnetic field strength and CSM density and
structure. Then as the shock wave overtakes more and more of the CSM,
the radio emission becomes progressively less absorbed, leading to a
characteristic frequency-dependent ``turn on'' first at higher
frequencies and later at lower ones.

Following the prescription of \citet{Chevalier1982b} the optically thin
radio luminosity of this shell can be described by

\begin{equation}
L_{\nu}\propto 4 \pi R^{2} \Delta R\ K\ B^{(\gamma+1)/2}\ \nu^{(\gamma-1)/2}\ e^{-\tau}
\end{equation}

\noindent where $L_{\nu}$ is the radio luminosity at frequency $\nu$,
$\Delta R$ is the thickness of the synchrotron emitting region at
radius $R$, $B$ is the strength of the magnetic field, the
distribution of accelerated particles is assumed to take a power-law
form of $N(E)=KE^{-\gamma}$, and $\tau$ is the absorption
opacity. Both $B$ and $K$ scale with the thermal pressure $P$, which
itself scales with the CSM density $\rho_{\rm CSM}$ of the synchrotron
emitting region. The opacity $\tau$ likewise scales with $\rho_{\rm
CSM}$, although in the case of FF absorption it is proportional to the
integrated density along the line of sight. We assume here that all
absorbing media are purely thermal, ionized hydrogen with opacities
proportional to $\nu^{-2.1}$. We contend that the FF opacity dominates
over the SSA opacity given that (1) the shock velocity required by
SSA, as implied by the radio luminosity and time of peak
emission \citep[see, e.g.,][and references therein]{Chevalier2006}, is
many times lower than either the optically inferred late-time shock
velocity or the expected early-time shock velocity, and (2) the
absorption does not deviate substantially from
$\tau \propto \nu^{-2.1}$ where it is possible to constrain it, as
expected for FF. By comparison, SSA should scale as $\tau_{\rm
SSA} \propto \nu^{\alpha-2.5}$, which is too
steep to fit any of our well-sampled epochs. In this case, the FF
opacity is given by

\begin{equation}
\tau_{\rm ff} \approx3.28\times10^{-7} \biggl({T_{e} \over 10^{4}\,{\rm K}}\biggr)^{-1.35} \biggl({\nu \over 1\ {\rm GHz}}\biggr)^{-2.1} \biggl({EM \over {\rm pc\,cm^{-6}}}\biggr)
\end{equation}

\noindent where the emission measure, EM, is given as

\begin{equation}
EM=\int^{\infty}_{r} N_{e}^{2} dr
= \int^{\infty}_{r} \biggl({\rho_{\rm CSM} \over \mu_{\rm H} m_{\rm
H}} \biggr)^{2} dr.
\end{equation}

\noindent Thus the unabsorbed radio luminosity $L_{\nu}$ and 
FF opacity $\tau_{\rm ff}$ afford us two completely independent
opportunities to trace the CSM density profile.

\begin{figure}
\vspace{-0.1in}
\centerline{
\hglue-0.1cm{\includegraphics[width=9.0cm]{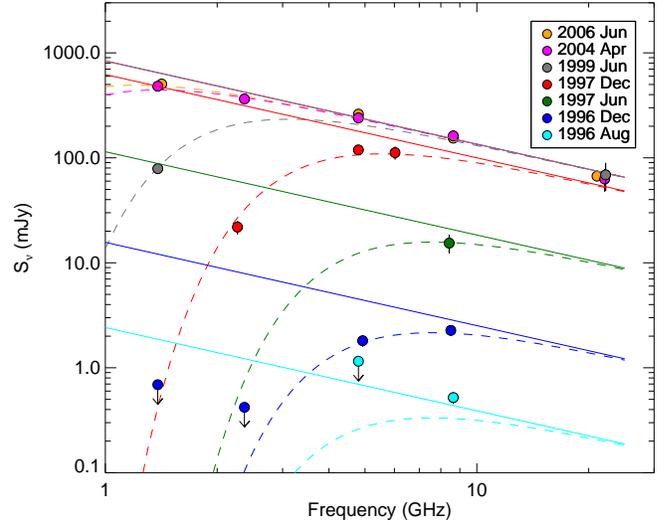}}
}
\vspace{-0.0in} 
\figcaption[sn1996cr_radiospec.eps]{
%
%
Radio spectra of \hbox{SN\,1996cr} over several epochs. Solid circles
denote data, while solid and dashed lines denote intrinsic and
absorbed radio spectra from our best-fitting model. We find dramatic
changes in the low-frequency slope indicative of time-varying
absorption, but relatively little change to the high-frequency slope
additionally suggesting a strong increase in the CSM density.
%
%
\label{fig:sn1996cr-radiospec}}
\vspace{0.2cm} 
\end{figure} 

With the above in mind, we now examine the radio data
for \hbox{SN\,1996cr}. Fig.~\ref{fig:sn1996cr-lc} demonstrates that
there are strong upper limits at four observable frequencies from the
earliest time of explosion up through the end of 1996. These
non-detections imply either strong early absorption or the presence of
a low-density cavity. Over the next year, we see a dramatic rise in
the radio emission, first at higher frequencies and later at lower
ones. The inverted rise, however, is atypical and cannot solely be the
result of dwindling absorption, as is common in other RSNe. To make
this point clearer, we present the radio spectrum of \hbox{SN\,1996cr}
for several epochs in Fig.~\ref{fig:sn1996cr-radiospec}. Much of
the radio data were taken separately, and thus a few of our adopted
late-time epochs actually span a several-month window; such an
approach is validated by the fact that strong, rapid radio variability
is not expected or observed at late times.

From Fig.~\ref{fig:sn1996cr-radiospec}, we see that the data taken
during the 2006 June and 2004 April epochs are only minimally absorbed
and thus provide a solid constraint both on the intrinsic synchrotron
spectral index ($\alpha=-0.79\pm0.02$) and the FF absorption beyond
the synchrotron-emitting region. Using our model radio spectra to
guide the eye (described below), we see the presence of increasing
absorption at the lowest frequencies as we progress back through the
1999, 1997, and 1996 epochs. This is a natural consequence of the
radio-emitting region overtaking the CSM which absorbs
it. Importantly, the observed spectral index between 4.8~GHz and
8.5~GHz does not change dramatically between 1996 and 2006. If we
assume that the intrinsic spectral index is constant, this implies
that the spectrum here is optically thin. The 8.5~GHz emission is thus
only modestly affected by FF absorption ($\sim20$\%) even at early
epochs and can be considered a relatively robust tracer of the
intrinsic radio luminosity at all times. As such, the 8.5~GHz flux
density appears to jump by a factor of $\approx$150 between 1996
August and 1997 December (515 days) and an additional factor of
$\approx$1.33 between 1997 December and 2004 April (2384 days). Given
that the radio-emitting region is spread over relatively large scales,
the radio emission should vary smoothly in time. Empirically, we find
that a broken power law, with indices $a_{1}=8.76\pm0.24$ and
$a_{2}=0.71\pm0.09$ between the three highly constrained epochs above,
provides an adequate fit to the data from other epochs aside from 1996
August ($\chi^{2}_{\nu}=2.65$), although there is clearly some
degeneracy with other parameters such as $\alpha$ and $\tau_{\rm ff}$
(see Fig.~\ref{fig:sn1996cr-radiospec}). This model should at least
provide a qualitative understanding of the light curve during
this overall period.

The early upper limits and dramatic increase in the radio luminosity
imply a sharp rise in $\rho_{\rm CSM}$ as well, highlighting the
possible transition from a fast, sparse stellar wind to a slow, dense
one. Such transitions typically lead to the formation of a wind-blown
bubble \citep[e.g.,][]{Weaver1977, Garcia-Segura1996a,
Garcia-Segura1996b}, and numerous researchers have explored the
subsequent interaction between the complex CSM associated with a
bubble and the SN blast wave \citep[e.g.,][]{Chevalier1989,
Tenorio-Tagle1990, Tenorio-Tagle1991, Dwarkadas2005}. Quantitative
constraints on the density are difficult to obtain, however, since the
radio luminosity cannot be related to the CSM density in the 
standard manner \citep[i.e., self-similar solutions such that
$L_{\nu}\propto \rho_{\rm CSM}^{0.4}$--$\rho_{\rm CSM}^{0.7}$ for
$\alpha$$=$0.79;][]{Chevalier1996} and requires hydrodynamical
simulations to account for the interaction properly.

Our constraints on $\tau_{\rm ff}$, on the other hand, are more
straightforward to calculate. Using the Levenberg-Marquardt
least-squares method we directly constrain $\tau_{0}=\tau_{\rm
ff}(\nu/{\rm 1\,GHz})^{2.1}$ to be 29.3$\pm$6.1, 15.8$\pm$1.1,
4.1$\pm$0.6, 0.69$\pm$0.10, and 0.57$\pm$0.11 for the 1996 December,
1997 December, 1999 June, 2004 April, and 2006 June epochs,
respectively.  Unfortunately, the 1996 August and 1997 June epochs
only have 8.5~GHz measurements and do not provide strong constraints,
although we are encouraged by the fact that the latter is consistent
with our adopted light-curve model. For the five epochs where we have
good spectral constraints, we solve for $\rho_{\rm CSM,shell}$ explicitly
in four contiguous shells such that
\begin{equation}
\rho_{\rm CSM, shell}\approx 3.49\times10^{3} \biggl({T_{e} \over 10^{4}\,{\rm K}}\biggr)^{1.35} \biggl({ \tau_{0}^{\rm in}- \tau_{0}^{\rm out} \over 
\int^{r^{\rm out}}_{r^{\rm in}} f(r) dr}\biggr)^{0.5} {\rm cm}^{-3}
\label{eq:rho_csm}
\end{equation}

\noindent where we have assumed $\mu_{\rm H}=0.6$ and $T_{e}=$10,000~K. 
The inner and outer radii $r^{\rm in}$ and $r^{\rm out}$, in pc, are
taken from $v_{s}t$;\footnote{While a self-similar solution is
unlikely to hold under our current physical conditions, it is
instructive to note that in such a scenario, the radii and velocity
would evolve as $r=(v_{s}/m) t$ and $v_{s}\propto t^{-m}$, where $m$
will range between 0.85$-$0.95 for most cases. This would lead to
estimated shell radii only $\sim20$\% larger and should highlight the
relative quality of our radial constraints.} we note that the
outermost radius is constrained to be $\approx2.8\times10^{17}$~cm on
2007-06-24 from our VLBI observation. The to-be-determined
$r$-dependence of $\rho_{\rm CSM, r}$ is encompassed by $f(r)$. The
form of $f(r)$ is iteratively determined by ensuring that it is
consistent with the slope obtained between adjacent shells of
$\rho_{\rm CSM, shell}$.

We next turn to the evolution of the shock velocity, $v_{s}$, which we
need in order to determine the shell radii.  From our optical
spectrum, the ejecta (as traced by the O lines) and broad H$\alpha$
line appear to be expanding at a variety of velocities between
$\approx$2000--5800 km~s$^{-1}$, with the symmetric red and
blueshifted velocities of \hbox{$\approx$3250--3550~km~s$^{-1}$}
standing out as the most likely values tracing the blast wave. We thus
adopt 3400~km~s$^{-1}$ for the current shock speed.  Clearly after the
shock impacts a high-density region, its velocity will be reduced
considerably. During its subsequent evolution in such a region,
however, its velocity should evolve strongly, as can be seen from
radio observations of SN\,1987A \citep{Manchester2002, Gaensler2007}
and simulations thereof \citep{Dwarkadas2007a}. Therefore, without any
further information, a first approximation is to assume a constant
velocity comparable to that derived from the optical spectrum.
Interestingly, the high X-ray temperature, if it is in thermal
equilibrium, also yields an equivalent velocity of
$\approx$3400~km~s$^{-1}$, although here the velocity is almost
certainly related to the reverse rather than forward shock and thus
may simply be coincidence.

We therefore adopt an evolution for the blast wave velocity as
follows. The shock began with a high initial speed which it maintained
for 1--2 yrs. By 1996 August it encountered a dense shell of
wind-swept material, and the shock velocity dropped rapidly by a
factor proportional to the square root of the density gradient it
encountered. This density gradient also initiated a rise in the radio
luminosity by a factor of $>200$. Beyond 1996 August, we assume the
velocity quickly arrived at a constant value of
$\sim$3400~km~s$^{-1}$. These assumptions allow us to determine
approximate radii of 0.054~pc, 0.055~pc, 0.059~pc, 0.064~pc, 0.081~pc,
and 0.089~pc for the 1996 August, 1996 December, 1997 December, 1999
June, 2004 April, and 2006 June epochs, respectively. 

If we take the radius on 1996 August as our shell impact date, then,
depending on the 380~day ambiguity in the explosion date, the average
initial shock velocity prior to this is constrained to be
$\sim$37,000--139,000~km~s$^{-1}$ (0.12$c$--0.46$c$). The lower value
is in line with theoretical and observational
expectations \citep[e.g.,][]{Chevalier2006}, while the upper value is
probably unrealistic. Such a high average shock speed would be
difficult to maintain for more than several days in a typical or even
underdense CSM environment. If the explosion did come later, then one
or more of our assumptions above may be incorrect. The late-time
velocity of \hbox{SN\,1996cr} could be larger than our adopted value:
the largest observed velocity in the optical spectrum is
$\approx$5,800~km~s$^{-1}$, for instance, and provides a plausible
upper limit to the current shock speed; adopting this value yields
initial velocity constraints a factor of $\approx$3
lower. Alternatively, the transition between initial and final
velocities could have been gradual, taking many weeks to years to
fully drop from one extreme to the other: the proposed CSM shell
of \hbox{SN\,1996cr} could have asymmetries akin to those seen
in \hbox{SN\,1987A}'s ring, for instance, which are likely responsible
for the onset of hotspots \citep[e.g.,][]{Sugerman2002}; a long-lived
velocity transition could again lower initial velocity constraints by
a factor of a few. Finally, the VLBI radius constraint could be too
large, due either to our adopted ring model or our adopted distance to
SN\,1996cr: we note that uncertainties in our adopted VLBI model are
estimated to be only $\sim$5--20\%, while our adopted distance could
vary by a factor of $\la$2.7 at most. Despite these potential
problems, though, it is encouraging that our overall velocity model,
assuming the lower initial shock velocity at least, is remarkably
similar to the actual time evolution of the shock velocity
in \hbox{SN\,1987A} as deduced from radio
observations \citep[e.g.,][]{Manchester2002}.  In SN\,1987A, the shock
expanded at $\sim$35,000~km~s$^{-1}$ for its initial three years,
before impacting with the H\,{\sc ii} region and slowing
to $\sim$4500~km~s$^{-1}$.

\begin{figure}
\vspace{-0.1cm}
\centerline{
\includegraphics[width=9.0cm]{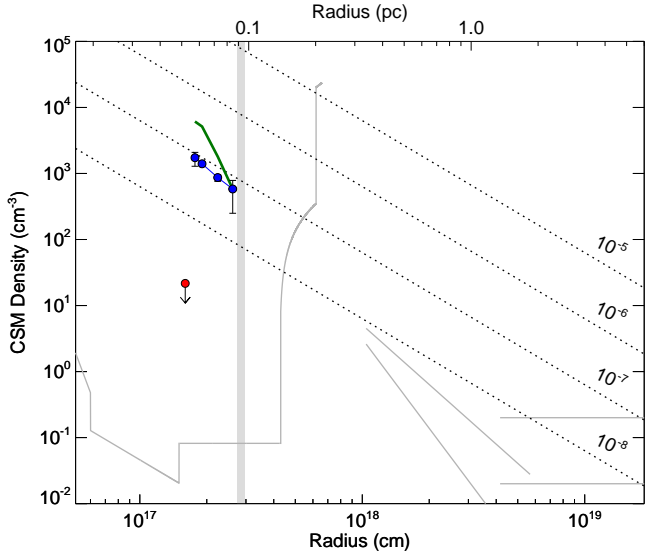}
}
\vspace{-0.0in} 
\figcaption[sn1996cr_csmdensity.eps]{
%
%
Estimated CSM density profile for \hbox{SN\,1996cr}; see
$\S$\ref{radio_ltcrv} for full details. Radii are determined assuming
a constant velocity of 3400~km~s$^1$ since 1996 December, anchored by
our VLBI size constraint shown as a thick grey line. The CSM densities
are estimated from the decreasing free-free absorption coefficient
assuming a fully-ionized CSM (see \ref{radio_ltcrv} for details), and
hence only trace the ionized portion of the CSM at these radii. The
green curve denotes the density profile obtained assuming $\rho_{\rm
CSM} \propto r^{-2}$, which is clearly inconsistent with this
assumption. The blue points and curve show the density profile when
modeled in a self-consistent manner, such that $\rho_{\rm CSM} \propto
r^{-s}$, where $s$ is the iteratively determined slope between
subsequent data points to be $-2.55$, $-2.93$, and $-2.88$,
respectively. We caution, however, that the exact slope $s$ depends
intimately upon our adopted velocity evolution, estimated radio size,
and adopted distance. The early upper limit (red) is the required
density jump needed to reconcile the average estimated initial shock
velocity with the much smaller average velocities inferred from the
optical spectrum, assuming the density represented by the last
measured point is fully ionized. The dotted lines denote constant
values of $\mdot/v_{\rm w}$ in units of
M$_{\odot}$~yr$^{-1}$~km$^{-1}$~s and bracket the typical values for
red supergiant ($10^{-5}$) and Wolf-Rayet ($10^{-8}$) winds. For
comparison, we also show the crude density profile (thin grey curves)
for \hbox{SN\,1987A}: the inner segment (left) is the density
structure required to explain both the X-ray and radio
emission \citep{Dwarkadas2007b}, while the outer lines (right) denote
the upper and lower bounds on the density structure as deduced by
light echoes \citep{Sugerman2005}.
\label{fig:sn1996cr-csmdensity}}
\vspace{0.0cm} 
\end{figure} 

Using Equation~\ref{eq:rho_csm}, we can now place constraints on
$\rho_{\rm CSM, r}$ in each shell of material overtaken between the
five epochs as shown in Fig.~\ref{fig:sn1996cr-csmdensity}. Since we
do not know how far out the remainder of the free-free absorbing
material extends into the CSM, we do not include the last $\tau_{0}$
value in this analysis. $H\alpha$ observations from the Southern
H-Alpha Sky Survey Atlas \citep{Gaustad2001} provide constraints on
the emission measure toward the Circinus Galaxy of
$\approx$10,000--25,000 pc~cm$^{-6}$, or equivalently
$\tau_{0}=(0.0033$--$0.0082)(T_{e}/10^{4}~K)^{-1.35}$. Thus for any
reasonable temperature of the warm interstellar medium (ISM), it
appears that much of the remaining free-free absorption from our last
measurement probably resides close to the SN in the CSM itself. Future
low-frequency radio measurements could constrain this potential
decline in the absorption. Oddly, we find that the $\rho_{\rm CSM,
r} \propto r^{-s}$ profile derived from the absorption, where
$s\approx2.5$--2.9, seems at odds with the flat late-time evolution of
the radio luminosity itself, which would naively imply
$s\approx1$.\footnote{Radiation losses or strong gradients in the
magnetic field could change $s$
substantially \citep[e.g.,][]{Fransson1998}. Note, however, than any
additional loss mechanism would have to leave the late-time spectrum
relatively unperturbed.}  The strong disconnect between the two may
indicate that a large portion of the CSM being overtaken is actually
neutral, and thus not traced by the radio absorption. Since
recombination is a strong function of density, this disconnect may
further imply that there is significant clumping within the CSM
of \hbox{SN\,1996cr}.

Although we are unable to constrain the density profile encountered by
the blast wave at early times via $\tau_{\rm ff}$, the early implied
velocity drop argues for a density increase of $\ga$120, while the
radio luminosity jump supports a similar increase (even larger if
$L_{\nu}\propto \rho_{\rm CSM}^{0.4}$--$\rho_{\rm CSM}^{0.7}$
holds). Thus it appears that we are seeing the transition between two
distinctly different density regions separated by a dense shell,
possibly associated with a wind-blown
bubble \citep[e.g.,][]{Dwarkadas2005}. If we assume the SN exploded
into a sparse stellar wind ($\rho_{\rm CSM} \propto r^{-2}$)
initially, then we might expect the shock to adhere to a self-similar
solution prior to 1996 August \citep[i.e., radio emission evolves as
$L_{\nu}\propto t^{-0.5}$--$t^{-1.5}$; CF94,][]{Chevalier1996}. Thus
for explosion dates earlier than 1995-06-07 (i.e., day 98 for
$t_{0}$$=$1995-02-28), our 8.5~GHz upper limit would constrain the CSM
density in this region to be even lower still if no further SSA or FF
absorption is present. A lower density, however, would imply a
substantially larger density jump later on, which is difficult to
reconcile with our current shock velocity evolution. This suggests
that either the SN went off after this date, the early density did not
follow a $r^{-2}$ profile, or there was additional heavy absorption
that affected even the 8.5~GHz data.

At the assumed radii, the CSM density of the shells equates to $\sim$
0.05~$M_{\odot}$ located between
(1.7--2.8)$\times10^{17}$~cm. Additionally, the SN blast wave took only
$\approx1$--2~yr to reach this high-density region, equating to
$\sim$(1--2)$v_{\rm s}/v_{\rm w}$~yr in the progenitor time frame. For
reasonable wind velocities, this implies the region was formed a mere
$\sim$70--14,000~yr prior to the explosion.

For comparison, the density for a constant progenitor wind (i.e.,
constant $\mdot/v_{\rm w}$) can be written as
\begin{equation}
\rho_{\rm CSM}=6.3\times10^{-12}
\biggl({\mdot \over {\rm M_{\odot}~yr^{-1}}}\biggr)
\biggl({v_{\rm w} \over {\rm km~s}^{-1}}\biggr)^{-1}
\biggl({r \over 10^{16}~{\rm cm}}\biggr)^{-2},
\end{equation}

\noindent and a few lines of constant $\mdot/v_{\rm w}$ are plotted in 
Fig.~\ref{fig:sn1996cr-csmdensity}. Notably, the evolution
of \hbox{SN\,1996cr} beyond $\sim$10$^{16}$~cm appears to be
consistent with a constant wind of $\mdot/v_{\rm
w}\sim10^{-7}$--$10^{-8}$, typical of a Wolf-Rayet (WR)
stage. However, it seems to have transitioned to a much sparser wind
of $\mdot/v_{\rm w}\la10^{-9}$ before this (i.e., during its last many
decades/centuries).

There are very few scenarios that can lead to such a CSM density
distribution, with an initial low density, followed by a large density
jump, and then a region of decreasing density. Such a distribution is
reminiscent, however, of a wind-blown bubble, wherein the interaction
of a star's stellar wind with either a previous stage of the stellar
wind or with a constant-density medium such as the ISM, results in
material being swept up into a thin dense shell between the two
regions. Going outwards in radius, the bubble density distribution is
described by \citep{Weaver1977, Dwarkadas2005}: A region of freely
expanding wind decreasing as $r^{-2}$, the wind termination shock, a
constant-density region of shocked wind, the thin dense shell of
swept-up ambient material and the outer medium, which could be another
wind region with density decreasing as $r^{-2}$.  The circumstellar
distribution derived from the radio observations seems to fit this
well, with the high-density jump denoting the location of the dense
shell, which we may not have seen from free-free absorption if the
dense shell was not highly ionized. Interior to the shell is the low
density shocked and unshocked wind, for which we only have upper
limits. Exterior to the shell is the unshocked ambient medium,
presumably also a wind, which we verify from our free-free
absorption. If true, this picture again indicates that the stellar
parameters changed significantly just a few years to a few thousand
years before the death of the star, and that this region was formed by
wind-wind interaction. Although the shell is thin, it is extremely
dense, and the shock colliding with this region could be slowed
dramatically, and lead to an increase in radio and X-ray emission, as
is observed.

SN\,1996cr can be compared to other observed young RSNe, among
which \hbox{SN\,1987A} \citep[see][for reviews]{McCray2005,
McCray2007} is again a good starting point because of its rising radio
light curve. \hbox{SN\,1987A} was detected at radio wavelengths when
it first went off, but this initial outburst was very
short-lived \citep{Turtle1987} and attributed to shock acceleration of
synchrotron-emitting electrons in the stellar wind close to the star
at the time of the explosion \citep{Storey1987, Chevalier1987}. Radio
emission was again detected from the supernova after $\sim$3~yr in
quiescence, at first rising in a dramatic fashion for several hundred
days and increasing more or less monotonically thereafter (see
Fig.~\ref{fig:sn1996cr-lc}). Unlike the early absorption in
\hbox{SN\,1996cr}, however, the observed spectral index of \hbox{SN\,1987A} 
has remained relatively constant between $-1.0$ to $-0.9$, indicating
unabsorbed synchrotron emission at all times during this
re-emergence \citep{Manchester2002}. The steady rise
in \hbox{SN\,1987A} is also much steeper than in \hbox{SN\,1996cr} and
$\ga10^{4}$ times fainter, indicating a markedly different CSM-density
distribution, and perhaps sharper density contrasts. The constant
monitoring at radio, optical, and X-ray wavelengths has illuminated
the density structure surrounding \hbox{SN\,1987A} and directly
confirmed multiple evolutionary stages of mass
loss \citep{Chevalier1995, Sugerman2005, Dwarkadas2007c,
Dwarkadas2007a}. Unfortunately such consistent monitoring is lacking
for \hbox{SN\,1996cr}, leaving us with only a crude understanding of
its density structure to date. Clearly the CSM distribution
around \hbox{SN\,1996cr} appears to be much denser and more compact
than \hbox{SN\,1987A}, in addition to potentially showing a pile-up of
material around $\sim10^{17}$~cm. We note that more modest structure
in the CSM density, perhaps from the wind-wind interaction of multiple
evolutionary stages or inhomogeneities within a particular stage, has
been observed in many well-sampled RSNe light curves now, including
\hbox{SN\,1978K} \citep{Smith2007a}, 
\hbox{SN\,1979C} \citep{Montes2000},
\hbox{SN\,1980K} \citep{Montes1998}, and 
\hbox{SN\,1993J} \citep{Bartel2002, Weiler2007}, and 
demonstrates that perturbations from $\rho_{\rm CSM} \propto r^{-2}$
are probably commonplace. With finer sampling in the future, we should
be able to place stronger constraints on the outlying density
structure in \hbox{SN\,1996cr}.

\subsection{X-ray Light Curve}\label{xray_ltcrv}

We turn next to the X-ray light curve of \hbox{SN\,1996cr}. The
Circinus Galaxy was observed with a variety of \hbox{X-ray}
instruments as shown in Fig.~\ref{fig:sn1996cr-lc} and
Table~\ref{tab:data_xray}. From these, we find that \hbox{SN\,1996cr}
exhibits strong, atypical temporal evolution. Unlike most SNe,
\hbox{SN\,1996cr} is not detected at early times despite relatively
frequent and sensitive observations. Since thermal X-ray emission effectively
traces the density of the CSM, these early \hbox{X-ray} upper limits,
especially when compared to the detections later on, again indicate
the presence of either strong early absorption ($N_{\rm
H}\ga2\times10^{22}$~cm$^{-2}$) or a low-density cavity immediately
surrounding the progenitor. Our radio constraints suggest in fact that
both are present and likely to affect the X-ray emission: the mass
associated with the FF absorption is already marginally enough to
obscure an extrapolation of the detected X-ray slope, while the lower
density implied by the radio luminosity jump would also result in
decreased early X-ray emission. When finally detected on 2000-01-16, the
\hbox{0.5--2} and \hbox{2--10~keV} \hbox{X-ray} fluxes are best-fitted
as $\propto$ $t^{1.02\pm0.17}$ and $t^{0.80\pm0.10}$,
respectively. The slopes are consistent with each other to within the
errors, suggesting that any apparent emerging soft component is
marginal and that the detected portion of the rise is likely to be
optically thin. This strong, continued rise is in stark contrast to
the evolution of nearly all other SNe detected to date, which are
theoretically expected to decline as $\propto$~$t^{-1}$--$t^{-0.4}$
(e.g., CF94) and observationally follow suit albeit with more
scatter \citep[e.g., see Fig. 2 of][]{Immler2005}. Sustained increases
in the \hbox{X-ray} have only been observed
for \hbox{SN\,1987A} \citep{Park2006}, \hbox{SN\,2006jc} \citep{Immler2008},
and perhaps marginally for \hbox{SN\,1978K} \citep[based upon a single
early upper limit;][]{Schlegel2004}.

The X-ray luminosity
of the forward and reverse shocks at a particular radius $r$ is given by

\begin{equation}
L_{\rm X}=4\pi \int \Lambda_{\rm ff}(T_{e,i})~n_{e,i}^{2}~r_{i}^{2}~dr,
\label{eq:lx}
\end{equation}

\noindent where the index $i$ refers to values for either the forward or reverse
shock and $\Lambda_{ff}$ is the temperature-dependent cooling rate
\citep{Chevalier2003}. The detection of strong He-like Fe and Si
emission lines in the X-ray spectra \citep{Bauer2001} suggests the
presence of heavy elements in the X-ray emitting material, and
therefore that the X-ray emission stems from the shocked ejecta
material found behind the reverse shock. 

Because the progenitor CSM distribution likely includes the
presence of pre-existing shocks and density discontinuities associated
with the wind-blown bubble (see $\S$\ref{radio_ltcrv}), the
interaction of the SN blast wave with this medium is complex. Due to
the presence of various discontinuities, the standard model
\citep[e.g][; CF94]{Chevalier1982b} of the interaction of a SN blast
wave with a power-law ambient density medium is not applicable, and we
must rely on detailed hydrodynamical models to infer the specific
mass-loss properties of the progenitor
\citep[e.g.,][]{Dwarkadas2005, Dwarkadas2007c, Dwarkadas2007b}. Furthermore 
the CSM may be at least partially photoionized, as is thought to be
the case for \hbox{SN~1987A} \citep[e.g.,][]{Chevalier1995}. As shown
in the previous section for a wind-blown bubble, even if the
progenitor star had a constant wind mass-loss rate and velocity (which
is not likely), the structure of the CSM can vary substantially with
radius. Thus we cannot assume that the CSM density directly translates
into the wind parameters. Nonetheless, we can make some qualitative
arguments based on Equation~\ref{eq:lx}.

Between days 1800--4400, we find that the intrinsic X-ray luminosity
scales almost linearly with time.  The radius over this time
period should increase by a factor of $\approx$1.6, depending on the
exact nature of the ejecta density profile and the CSM. The width of
the emitting region can be approximated to be about 0.1 times the
radius (e.g., CF94), resulting in an overall volume change by a factor
of $\la$5. From the X-ray spectrum, we do not see any significant
changes in the temperature or absorption, indicating that 
$\Lambda_{ff}(T_{i})$ remains approximately constant. If the X-ray
emission was arising in the shocked ambient medium, the density in
Equation~\ref{eq:lx} must decrease by a factor of $\approx$2.2, or
equivalently be $\propto r^{-1}$ to produce the observed
dependence. This is significantly different from either a constant
density or the $\approx r^{-2}$ dependence expected for a stellar wind
with constant parameters. Furthermore, for a density profile
decreasing as $r^{-1}$ the velocity would be expected to decrease with
time, and therefore the temperature (which is proportional to $v^2$)
would drop accordingly. The fact that this is not seen makes it
unlikely that the X-ray emission is arising from the shocked wind
(see Fig.~\ref{fig:sn_diagram}).

We instead contend that the emission is arising from shocked ejecta
behind (in a Lagrangian sense) the reverse shock. This scenario is
strengthened by the presence of heavy element emission lines in the
X-ray spectrum \citep[cf. Fig.~6 of][]{Bauer2001}, since such elements
are far more likely to reside in the processed ejecta than in the
progenitor's stellar wind. If the SN ejecta did collide with a dense
shell of CSM from a pre-existing wind-blown bubble, then the forward
shock would advance quite slowly, and the bulk of the X-ray emission
would arise from the reverse-shocked material. The post-shock
temperature just behind the reverse shock could be quite high
($\ga$$10^{9}$~K) if the density were low --- technically the density
could drop to zero for expansion in a constant-density medium.  The
electron temperature, however, might be an order of magnitude lower if
the electrons and ions have not yet reached equilibrium. The
temperature would not change appreciably over this period, consistent
with our observational constraints. The velocity of
$\approx$3400~km~s$^{-1}$ inferred from the X-ray temperature in this
case would refer to the electron temperature in the reverse shock, and
thus the actual shock velocity, as traced by the ion temperature,
could be considerably higher. Note that a similar linearly increasing
light curve profile was found for the hard X-ray emission from
SN\,1987A \citep{Park2005, Park2006}. \citet{Dwarkadas2007c} has
carried out numerical simulations that suggest that this emission
arises from the reverse shocked ejecta, as had been postulated
by \citet{Park2006}. One puzzle which remains, though, is how the
strong the H$\alpha$ and Oxygen emission-line luminosities are
powered. This discrepancy could be reconciled if there were holes in
the CDS, such that a small fraction of the X-ray emission could escape
while the rest went into exciting the strong optical line emission.

We can again gain some further insight by comparing \hbox{SN~1996cr}
to other SNe.  As stated above, \hbox{SN~1987A} has the most
similarities. Like \hbox{SN~1996cr}, \hbox{SN~1987A} was undetected at
X-ray wavelengths when it first went off, thereby excluding an
explosion directly into a slow, dense wind that is characteristic of a
red supergiant. Notably, \hbox{SN~1987A} was detected 130 days after explosion
\citep{Dotani1987, Sunyaev1987}, although the X-ray emission here was
attributed solely to radioactivity and appeared to rise and fall over
the following year. After this brief period, the X-ray emission went
undetected for another $\approx$~3~yrs before resurfacing above the
noise \citep{Hasinger1996}. Since this time, the X-ray emission has
continued to rise as shown in Fig.~\ref{fig:sn1996cr-lc}, with its
light curve well-modeled by a broken power law \citep{Park2006}.  
The reappearance of the emission at about 3 years was postulated as
due to the interaction of the SN blast wave with a region of ionized
wind material from the progenitor \citep{Chevalier1995}. They
suggested that this region is material photoionized from the dense
circumstellar shell that surrounds \hbox{SN~1987A}, formed as a result
of the interaction of the progenitor blue supergiant wind with a
pre-existing red supergiant wind.  Although the evolution of the star
may be different, a physical scenario broadly like that of
\hbox{SN~1987A}, with a region formed by mass-loss from a  progenitor star, 
 certainly seems plausible for \hbox{SN~1996cr}, although the
contrasting light-curve slopes, luminosities, and radial distances
clearly highlight the different CSM density structure. The only other
SN to exhibit a rise at X-ray wavelengths is \hbox{SN\,2006aj}, a
peculiar type\,Ib. The light curve of \hbox{SN\,2006aj} has been
ascribed to the interaction of the blast wave with a dense shell of
material left over from an LBV-like outburst of the SN progenitor
$\sim$2 years prior to the explosion \citep{Immler2008}.

The X-ray spectrum for \hbox{SN~1996cr}, moreover, is modeled by a hot
thermal plasma ($kT=13.4$~keV) similar to the early phase of \hbox{SN\,1993J}
\citep{Zimmermann2003} but substantially hotter than \hbox{SN~1987A}
($kT\approx3$~keV with an emerging $0.3$~keV component), and implies
that the conditions within the X-ray-emitting region are somewhat
different. The hotter temperature could indicate a higher shock
velocity, or a larger level of equilibration between the electrons and
ions. The X-ray spectrum of \hbox{SN~1996cr} is additionally absorbed
below $\sim$2~keV. While some fraction of the substantial non-Galactic
column density could be due to self-absorption, we consider this to be
minimal given that other \hbox{X-ray} point sources in the Circinus
Galaxy have similar \hbox{X-ray} derived column
densities \citep{Bauer2001}.

The rapid evolution we see here for \hbox{SN\,1996cr} has been noted
in a few other objects, such as \hbox{SN\,2002kg} \citep{VanDyk2006,
Maund2006} and \hbox{SN\,2006jc} \citep{Foley2007,
Pastorello2007}. Such an evolution has also been invoked to explain
the enormous energy output from \hbox{SN\,2006gy} \citep{Ofek2007,
Smith2007b}, although the lack of obvious CSM-interaction measures at
X-ray, H$\alpha$, and radio wavelengths here implies that perhaps
something more extreme is going on \citep[e.g.,][]{Smith2007c}.
Unfortunately, for many of the older, classic X-ray
emitting \hbox{type~IIn} SNe such as
\hbox{SN\,1978K} \citep{Schlegel2004, Smith2007a}, \hbox{SN\,1979C}
\citep{Immler2005}, and \hbox{SN\,1986J} \citep{Temple2005}, 
early-time behavior of the X-ray emission remains completely unknown
and thus the frequency and duration of sudden evolutionary changes in
SNe progenitors is still an open question. The X-ray evolution
of \hbox{SN\,1996cr} may allow us to explain the relatively flat
late-time X-ray light curves for objects like \hbox{SN\,1978K} and 
\hbox{SN\,1979C}, for instance.

\section{Conclusions}\label{conclude}

We have confirmed \hbox{SN\,1996cr} as one of the nearest, and
subsequently X-ray and radio-brightest, SNe to date. Archival optical
imaging data have constrained the explosion date to within
$\approx$1~yr prior to 1996-03-16, while a new VLT spectrum has
identified it as a type~IIn SN. The multi-wavelength constraints
detailed above point to a rich progenitor history. As is the case for
most type~IIn SNe, \hbox{SN\,1996cr} appears to have originated from a
massive star that shed its outer layers at some late evolutionary
stage, which now act as the target for the outgoing SN shock.  

The optical light curve hints at a mild re-brightening at late times,
possibly driven by line emission, or by X-ray photons being
downscattered and being emitted as optical photons.
From the recent optical spectrum, the strong narrow H emission points
to a significant accumulation of mass in the CSM prior to the SN. We
are additionally able to resolve the complex O emission into a number
of distinct high-velocity components with
FWHM$\sim$2000--3000~km~s$^{-1}$. The origin of such distinct O
complexes remains somewhat unclear, perhaps arising from one or more
of the following: structural asymmetries, differential absorption,
or multiple concentric shocks.

The \hbox{X-ray} and radio emission allow us to quantify the CSM
density, tracing out two apparently distinct regions, which cannot be
reconciled with a single constant-wind model. From the radio and X-ray
upper limits, we infer that the inner region 
($r\la1\times10^{17}$~cm) is likely quite sparse.  The outer region
[$r\approx(1.5$--$2.8)\times10^{17}$~cm], as traced by the early radio
absorption, radio luminosity jump, and X-ray emission, is not well
constrained, and is likely associated with the interaction region
between two distinct progenitor winds, which culminates in a thin,
dense wind-swept shell. This second region appears relatively dense
($\rho_{\rm CSM}\approx10^{3}$~cm$^{-3}$), potentially consistent with
a slow-velocity WR-like wind, although the CSM density profile traced
by the radio absorption ($\rho_{\rm CSM}\propto r^{-2.5}$) naively
seems at odds with the implied CSM density profiles from the radio and
X-ray luminosities ($\rho_{\rm CSM}\propto $constant or $r^{-1}$,
respectively). A more detailed understanding of the
CSM awaits sophisticated hydrodynamical models.

The most plausible explanation is one where the progenitor changed
evolutionary states just prior to explosion, likely transitioning from
a WR-like wind to an even faster, less dense wind, resulting in
portions of the previous CSM being swept-up to form a wind-blown
cavity surrounded by a dense shell. Such a scenario has been recently
proposed for the peculiar \hbox{type~Ib} \hbox{SN\,2006aj} \citep{Foley2007,
Pastorello2007, Immler2008}.  The radio and X-ray emission are
constant or still rising, demonstrating that the blast wave is still
making its way through the dense, swept-up shell at the edge of the
cavity.  The collision of the blast wave with the shell would result
in a transmitted shock expanding into the shell, and a reflected shock
expanding back into the SN ejecta. This reflected shock should be
distinguished initially from the reverse shock of the SN expanding
into the ejecta, although it will eventually overtake the reverse
shock. The overall effect could be one of several shocks and
rarefaction waves travelling back and forth through the ejecta. A
spectrum would reveal several different velocity structures, perhaps
as we observe in the O lines. The reason why it is seen mainly in O is
unclear, but it could indicate that this is an O-rich remnant, similar
to Cas\,A for example \citep[see ][and references
therein]{Fesen2001}. Or perhaps it is just that the reverse shock has
not penetrated very far into the ejecta, so that only the O-rich
layers have been shocked. We note that if the shock is indeed
progressing into shell of constant density, then the resulting density
gradients would not lead to the formation of a CDS at all \citep[e.g.,
see Fig. 1--4 of][]{Chevalier1982a}. If so, then the observed O
velocity components may instead arise from a disk or ring prehaps
similar to SN\,1987A.

Once the blast wave makes its way through the dense shell, it will
emerge from the other side, presumably as a weak shock. This should
encounter a relatively pristine wind and start to decline in both the
radio and X-ray bands.  Further regular monitoring
of \hbox{SN\,1996cr} should reveal this.  Once this happens, we will
be able to strengthen many of our current constraints and
assumptions.

We have compared \hbox{SN~1996cr} to the late-time behavior of
\hbox{SN~1987A}, which is the only other SN with a documented dramatic
increase at \hbox{X-ray} and radio wavelengths. \hbox{SN~1996cr} is
likely propagating through its CSM in a manner similar to
\hbox{SN~1987A} expanding into its \ion{H}{2} region and circumstellar ring
\citep[e.g.,][]{Park2006}. The \hbox{X-ray} and radio luminosities of
\hbox{SN~1996cr}, however, are several orders of magnitude brighter
and comparable to typical type~IIn SNe such as \hbox{SN~1978K} and
\hbox{SN~1979C}. This is presumably because the swept-up CSM is more compact, 
and perhaps, unlike \hbox{SN~1987A}, there is no intervening high-density
material encountered which slows down the ejecta until they reach a
dense shell, so the density jump is higher and subsequently more
dramatic. As we follow \hbox{SN~1996cr} over the next few decades, we
anticipate that it will transition into an evolution similar to other
SNe. It is intriguing to think that \hbox{SN~1996cr} might represent a
sort of ``bridge'' object between the extremes of
\hbox{SN~1987A} and these more luminous type~II SNe.
Although plagued by small number statistics, the fact that two
(although perhaps more if we count \hbox{SN\,1978K}
and \hbox{SN\,1979C}) of the $\sim$5--10 closest SNe to have exploded
in the past four decades show evidence for wind-blown bubbles implies
that the phenomenon is common, and may indicate that current SNe
searches are somehow biased against them at larger
distances. Systematic constraints exist only for type~Ibc
SNe \citep{Soderberg2006c, Soderberg2006a}, where late-time radio
observations indicate that only $\sim$2\% of initially undetected
targets show late-time detections, while $\sim$40\% of detections
exhibit abrupt light-curve variations.

The large mass associated with the pre-SN CSM surrounding
\hbox{SN\,1996cr} could have been the result of explosive mass loss
thousands of years before the SN
explosion \citep[e.g.,][]{Woosley2002, Heger2002}, qualitatively
similar to \hbox{SN\,2006aj} \citep{Foley2007, Pastorello2007,
Immler2008}, although the limited \hbox{X-ray} and radio data point to
a relatively smooth density profile, which implies a more gradual
process was at work.  Numerous other SNe
including \hbox{SN\,1994W} \citep{Chugai2004},
\hbox{SN\,2001em} \citep{Chugai2006, Bietenholz2007}, \hbox{SN\,2006gy}
\citep{Ofek2007, Smith2007b}, and \hbox{SN\,2006jc} \citep{Foley2007, Pastorello2007, Immler2008} 
all show signs of dense, enriched CSM likely produced by giant
mass-loss events just prior to their SNe, implying that this form of
mass-loss is relatively common. LBVs have often been invoked to
account for such episodes of extreme mass-loss
\citep[e.g.,][]{Gal-Yam2007, Smith2007b}. Indeed a giant outburst was seen 
in the case of  \hbox{SN\,2006jc} just 2 years prior to core-collapse
\citep{Pastorello2007}. Again it is unclear whether the explosive
nature of LBVs can be reconciled with the relatively smooth
mass-loss/density gradients observed for \hbox{SN\,1996cr}.

Our analyses clearly demonstrate that \hbox{SN\,1996cr} is a unique and
compelling target. Its proximity and flux enable a wealth of
observations that would be otherwise impractical for typical SNe. We
have reported on initial findings of a VLBI campaign underway to
resolve spatially \hbox{SN\,1996cr} at several radio wavelengths,
eventually enabling studies of its morphology and temporal
expansion. This will provide a more coherent picture, allowing us to
set strong limits on the expansion velocity from the resolved size and
potentially confirm the implied asymmetries we infer from the optical
spectrum. Follow-up {\it HST} imaging will provide constraints on any
potential light echoes in the vicinity of \hbox{SN\,1996cr}, possibly
allowing refined estimates of its explosion date, distance, and
outlying CSM structure. Likewise, follow-up \hbox{X-ray} observations
will hopefully elucidate the nature of the strong continued rise in
the \hbox{X-ray} band and better constrain properties of the reverse
shock region. Notably, \hbox{SN\,1996cr} is bright enough to be
efficiently monitored using the {\it Chandra} HETGS, enabling
high-resolution studies of its numerous \hbox{X-ray} emission lines
and their potential evolution; an in-depth discussion of the
current \hbox{X-ray} spectral constraints, including a composite {\it
Chandra} HETGS spectrum from the data presented here, are forthcoming
(F. Bauer et al., in preparation). Additionally, future optical
spectroscopic studies should allow investigations into the temporal
behavior of complex emission lines found in our discovery
spectrum. Infrared imaging and spectroscopy will help address the
potential for dust formation; notably \hbox{type~II} SNe are thought
to contribute strongly to the overall dust content of galaxies.
Looking ahead to the near future, if \hbox{SN\,1996cr} remains
relatively bright at millimeter wavelengths, we anticipate that
observations with ALMA will provide a substantial leap in terms of
understanding the structure and composition of \hbox{SN\,1996cr} and
its relation to both \hbox{SN\,1987A} and more typical \hbox{type~II}
SNe.

If any researchers have further serendipitous multi-wavelength
observations of \hbox{SN\,1996cr} during the critical 1995--1996 period, we
would be grateful to learn of these.

\acknowledgements
We thank Bjorn Emonts and Bryan Gaensler for help in reducing archival
ATCA observations, Roger Chevalier, Claes Frannson, Dick McCray, Kurt
Weiler, and Christopher Stockdale for useful discussions about the
nature of the multi-wavelength emission, Baerbel Koribalski for
increasing our awareness of Australian online archives and access to
proprietary ATCA observations, Ernesto Oliva and Alessandro Marconi
for access to reduced NTT data, Bruno Leibundgut and Rob Fesen for
their spectra of \hbox{SN\,1986J} and \hbox{SN\,1979C}, Philip
Edwards, Steven Tingay, and Tasso Tzioumis for support of the VLBI
observations, Kevin Hurley for directing us to revised IPN results
which allowed us to reject an otherwise tentative GRB identification,
John Raymond for helping to improve our emission-line tables, and
the anonymous referee for useful comments that improved the content
and presentation of the paper.
We gratefully acknowledge the financial support of Chandra
Postdoctoral Fellowship Award PF4-50032 (FEB), NSF award AST-0319261
(VVD), NASA STScI grant HST-AR-10649 (VVD), NASA LTSA grant
NAG5-13035 (WNB), the Leverhulme Trust (SJS) and the ESF EURYI
scheme (SJS).

{\it Facilities:} 
\facility{ATCA},
\facility{ASCA (GIS, SIS)}, 
\facility{AAT (TAURUS)}, 
\facility{BeppoSAX (LECS, MECS)}, 
\facility{CGRO (BATSE)},
\facility{CXO (ACIS, HETGS)},
\facility{Hobart}, 
\facility{HST (WFPC2, NICMOS)}, 
\facility{Max Planck:2.2m (IRAC2)}
\facility{Mopra}, 
\facility{NTT (SUSI)}, 
\facility{Parkes}
\facility{ROSAT (HRI)},
\facility{Swift (UVOT, XRT)},
\facility{UKST}, 
\facility{VLT (FORS)}, 
\facility{XMM (p-n, MOS)},

\end{document}